# Multipath cycleGAN for harmonization of paired and unpaired low-dose lung computed tomography reconstruction kernels

Aravind R. Krishnan[1], Thomas Z. Li[2], Lucas W. Remedios[3], Michael E. Kim[3], Chenyu Gao[1], Gaurav Rudravaram[1], Elyssa M. McMaster[1], Adam M. Saunders[1], Shunxing Bao[1], Kaiwen Xu[4], Lianrui Zuo[1], Kim L. Sandler[5], Fabien Maldonado[6,7], Yuankai Huo[1,3], Bennett A. Landman[1,2,3,5,8]

[1]Department of Electrical and Computer Engineering, Vanderbilt University, Nashville, TN, USA, [2]Department of Biomedical Engineering, Vanderbilt University, Nashville, TN, USA, [3]Department of Computer Science, Vanderbilt University, Nashville, TN, USA, [4]Insitro, South San Francisco, CA, USA, [5]Department of Radiology and Radiological Sciences, Vanderbilt University Medical Center, Nashville, TN, USA, [6]Department of Medicine, Vanderbilt University Medical Center, Nashville, TN, USA, [7]Department of Thoracic Surgery, Vanderbilt University Medical Center, Nashville, TN, USA, [8]Vanderbilt University Institute of Imaging Science, Vanderbilt University Medical Center, Nashville, TN, USA

**Abstract**

**Background:** Reconstruction kernels in computed tomography (CT) reconstruction introduce variability in the spatial resolution and noise distribution that creates systematic differences in quantitative imaging measurements, e.g. emphysema characterization in lung imaging. Therefore, an appropriate choice of reconstruction kernel is necessary for obtaining consistent measurements during quantitative image analysis.

**Purpose:** We explore training a harmonization model based on different types of data with paired and unpaired CT reconstruction kernels in a low-dose lung cancer screening cohort and validate our approach through quantitative CT measurements.

**Methods:** We develop a multipath cycleGAN model that enables kernel harmonization by leveraging a shared latent space, domain-specific encoder-decoder architectures and discriminators trained using a mixture of paired and unpaired data. To evaluate our approach, we train our model across 42 different combinations of reconstruction kernels on 100 scans each from seven representative kernels obtained from the National Lung Screening Trial (NLST) dataset. We harmonize 240 withheld scans from each representative kernel to the style of a reference soft kernel and evaluate percent emphysema quantification before and after harmonization, followed by a general linear model analysis that highlights the impact of age, sex, current smoking status, and kernel on emphysema quantification. We also evaluate percent emphysema before and after harmonization for all the soft kernels harmonized to a reference hard kernel. Furthermore, we quantify anatomical consistency in unpaired kernels by comparing segmentations of lung vessels, muscle and subcutaneous adipose tissue obtained from TotalSegmentator between the non-harmonized and harmonized images. We compare the performance of our model to the traditional cycleGAN and the switchable cycleGAN model which uses a single generator with split adaptive instance normalization.

**Results:** The proposed multipath approach mitigates differences in percent emphysema scores for the paired kernels as observed from the median root mean square error (RMSE) and 95% limits of agreement obtained from Bland Altman plots ($p<0.05$). For the unpaired kernels harmonization mitigates confounding differences in emphysema measurement ($p>0.05$). Additionally, consistently high Dice of anatomical segmentation before and after harmonization demonstrate that muscle and subcutaneous adipose tissue anatomy was conserved in the unpaired kernels. On the lung vessels, the proposed approach shows reasonable overlap as evidenced from the Dice scores and effect size.

**Conclusions:** Paired and unpaired kernel harmonization with a shared latent space multipath cycleGAN mitigates errors in emphysema quantification and maintains anatomical consistency after harmonization.

## 1. Introduction

In computed tomography (CT), there exists a trade-off spectrum between spatial resolution and pixel noise: scans reconstructed with a "hard" kernel have higher spatial resolution and increased pixel noise while those reconstructed with a "soft" kernel have lower spatial resolution and decreased pixel noise[1] (**Figure 1**). Hard kernels are beneficial when lung or bone is the anatomy of interest whereas soft kernels are beneficial for soft tissue like the mediastinum[2]. However, the sharpness of the kernel impacts the strength of distinct spatial frequencies and image features; these changes result in inconsistent quantitative measurements across kernels [3]. In thoracic imaging, emphysema quantification[4,5], body composition assessment[6], coronary artery calcification[7] and radiomic feature reproducibility[8,9] are sensitive to the choice of kernel, making it difficult to compare scans obtained using different acquisition protocols in multi-centre and longitudinal studies.

Therefore, it is necessary to harmonize images reconstructed with different kernels to ensure consistency and reliability in quantitative analysis.

Kernel harmonization is an approach that mitigates systematic differences in quantitative measurements due to the variability of reconstruction kernels. The goal of kernel harmonization is to standardize the pixel noise in images reconstructed with different kernels to a common reference kernel, ensuring comparable and consistent quantitative measures. Progress has been made to implement kernel harmonization through physics-based and deep learning approaches. Physics based approaches generally involve the use of the point spread function (PSF) and modulation transfer function (MTF) that are modelled in the harmonizer. For instance, Ohkubo et al.[10] performed kernel filtering between two kernels using the ratio of the MTFs derived from the PSF. The residual intensity obtained on the difference images for lung cancer screening images and phantoms were mitigated after harmonization. Similarly, Sotoudeh-Paima et al.[11] developed CT HARMONICA, a physics-based harmonization method that harmonized CT images to isotropic resolutions of 0.75×0.75×0.75 mm and a global noise index described by Christianson et al.[12], to facilitate accurate emphysema quantification in phantoms and clinical data. By incorporating the MTF and an adaptive Wiener filter in their harmonizer, differences in emphysema-based measurements were mitigated. Zarei et al.[13,14] developed a harmonizer that incorporated the MTF into a deep neural network for harmonizing phantom images. Their approach mitigated differences in emphysema measurements, improved the detectability index of tumours in clinical data and mitigated the standard deviation of noise that is introduced by the kernel. While physics-based approaches are beneficial for kernel harmonization, the MTF and PSF need to be empirically estimated from phantoms which may not be available in retrospective studies.

Deep learning-based methods for kernel harmonization have shown potential for intra-vendor and cross vendor harmonization using a) supervised and b) unsupervised learning. Supervised learning methods rely on "paired" reconstruction kernels, meaning images reconstructed with hard and soft kernels of the same subject having a one-to-one pixel correspondence. In contrast, unsupervised methods use "unpaired" reconstruction kernels obtained from different vendors where there exists a difference in the underlying anatomical alignment and acquisition protocol. Lee et al.[15] developed a convolutional neural network (CNN) model inspired by Kim et al.[16] that learned to predict a residual map between a source kernel and target kernel for different pairs of reconstruction kernels obtained from a Siemens manufacturer. The residual learning strategy effectively captured differences between kernels, enabling standardization of emphysema measurements without the need for projection data. Eun et al.[17] leveraged super resolution networks and developed a squeeze excitation block network that learned to translate between reconstruction kernels obtained from a single vendor, showing improvements in image similarity metrics after harmonization. Inspired by the residual learning strategy, Choe et al.[18] leveraged the CNN model for kernel harmonization and highlighted its importance in reproducing radiomic features for different kinds of pulmonary nodules. Jin et al.[19] implemented a two-stage approach involving a truncation restoration network that compensates for artefacts followed by a kernel harmonization network for paired kernels obtained from four different manufacturers, showing considerable improvements in emphysema quantification.

Within deep learning methods, image-to-image translation is another approach for kernel harmonization that aims to learn a mapping between a source image and a target image, changing the style of the image while maintaining the contents[20]. Generative adversarial networks (such as pix2pix[21] for paired data and cycleGAN[22] for unpaired data) have gained popularity for kernel harmonization. One such approach was implemented by Selim et al.[23], where a novel dynamic window-based training was implemented using a generative adversarial network. The trained model reproduced radiomic features and showed good performance on image similarity metrics. Tanabe et al.[24] developed a pix2pix model that incorporated a region wise learning strategy on paired hard and soft kernels, minimizing differences in measurements for emphysema, body composition and coronary calcium. Similarly, in our previous work[25] incorporated the residual learning strategy into a pix2pix model and performed kernel harmonization on paired kernels obtained from a multi-vendor setting for lung cancer screening. Their approach mitigated differences in emphysema and body composition measurements and further highlighted reproducibility of radiomic features after harmonization. While CNNs and pix2pix models are reliable for kernel harmonization, they are limited to paired data from a given vendor, requiring multiple models for different pairs.

Unpaired kernel harmonization is an active area of research that poses challenges that include differences in image acquisition protocols and differences in the anatomical alignment of subjects across different manufacturers. Recent progress has facilitated cross vendor harmonization across unpaired kernels. For instance, Yang et al.[26] demonstrated kernel harmonization using a novel switchable cycleGAN model for head and facial bone kernels. Similarly, Selim et al.[27] developed a cross-vendor harmonization model using a cycleGAN that involved a self-attention module in the generator followed by a feature-based domain loss that helped in downstream radiomic feature assessment. Their approach improved the reproducibility of radiomic features with improved image similarity metrics. Gravina et al.[28] leveraged a cycleGAN that performed kernel harmonization on two kernels obtained for PET data, with their approach improving performance on image-based similarity metrics.

Existing approaches investigate paired and unpaired CT kernel harmonization by learning representational spaces that pertain to specific pairs. We propose that kernels obtained from various vendors can be jointly represented in a single latent space. Herein, we investigate paired and unpaired CT kernel harmonization across different combinations of reconstruction kernels in a low-dose lung cancer screening dataset with images obtained from multiple vendors using a proposed multipath cycleGAN trained in two stages. The key idea behind the multipath cycleGAN lies in the use of domain specific encoder-decoder architectures, stitched together through a shared latent space. We hypothesize that a shared latent space will:

- Enforce better consistency in emphysema measurements for paired reconstruction kernels as compared to separate latent spaces such as traditional cycleGAN[22].
- Mitigate kernel effects in emphysema quantification when all kernels are harmonized to a reference soft kernel, achieving better consistency compared to separate latent spaces for every domain.
- Mitigate kernel effects in emphysema quantification when all soft kernels are harmonized to a reference hard kernel.
- Enforce anatomical consistency for segmentations on the harmonized images for all paths that involve unpaired kernel harmonization to a reference soft kernel.

We compare our model with existing approaches that perform kernel harmonization that include the traditional cycleGAN[22] model that uses two separate latent spaces for image-to-image translation and the switchable cycleGAN model implemented by Yang et al.[26] that involves a cycleGAN model with a single generator. These baseline models are trained for all the paired reconstruction kernels and all paths that involve harmonization to a reference soft kernel and all soft kernels to a reference hard kernel. The performance of the baselines are compared with our multipath cycleGAN model for the following tasks: a) emphysema quantification on paired kernels b) emphysema quantification of all kernels harmonized to a reference soft kernel c) Impact of kernel on emphysema measurements using a general linear model and d) anatomical consistency on the harmonized images for the unpaired kernels using TotalSegmentator[29].

## 2. Methods

Inspired by prior research on paired and unpaired kernel harmonization, we leverage the cycleGAN model to develop a multipath cycleGAN model that harmonizes paired and unpaired kernels in a low dose lung cancer screening population from multiple vendors through a shared latent space trained in two stages. In our previous work[30], we developed a multipath cycleGAN model to address paired and unpaired CT kernel harmonization for images reconstructed from Siemens and GE scanners. This approach introduced a shared latent space that allowed for harmonization across different combinations of reconstruction kernels using domain specific encoder-decoder architectures and discriminators.

We expand upon our previous work by proposing a two-stage multipath cycleGAN model that involves a larger dataset from various kernels and standardized field of view across all images. In Stage one, we harmonize across all possible combinations of reconstruction kernels obtained from the Siemens and GE vendors. We incorporate an additional loss function between the input source kernel and output target kernel for a given path to ensure that the radio-opacity of the images do not shift during harmonization. In Stage two, we freeze the encoders and decoders trained in Stage one and train all discriminators to facilitate harmonization across all combinations of kernels from the Philips and GE vendors. We develop a strategy to select the best performance of the model on the validation data using emphysema scores and evaluate the model's ability to mitigate differences in emphysema scores for the paired and unpaired kernels. Additionally, we study the impact of age, sex, current smoking status and kernel on all the kernels that were harmonized to a reference soft kernel using a general linear model. We quantify anatomical consistency between the input source kernel and harmonized images across for all the unpaired reconstruction kernel paths using Dice scores obtained on skeletal muscle, lung vessels and subcutaneous adipose tissue.

## 2.1 Data selection

We harmonize CT images reconstructed using kernels from the National Lung Screening Trial (NLST), a randomized control trial that screened patients for lung cancer using low-dose CT (LDCT) scans and chest radiography scans[31]. Participants involved in this study were in the age range of 55-74 years, had a smoking history of 30 pack years or more and had last smoked within 15 years[31]. Each participant underwent three screenings, labelled T0, T1 and T2 at one-year intervals, with T0 considered as the baseline. We consider LDCT scans from the baseline that were reconstructed with different types of reconstruction kernels from multiple vendors. In a specific vendor, we choose subjects having a scan reconstructed using both a "hard" kernel and a "soft" kernel, following the NLST protocol[32], as "paired" reconstructions, where a one-to-one pixel correspondence is present between the scans. Across two vendors, there exists no overlap between subjects, resulting in "unpaired" reconstructions that lack a one-to-one pixel correspondence (**Figure 1**). We identify the following representative reconstruction kernel pairs: B50f (hard) and B30f (soft) from the Siemens vendor, D (hard) and C (soft) from the Philips vendor, BONE (hard), LUNG (hard) and STANDARD (soft) from the GE vendor. Subjects having scans reconstructed with the STANDARD kernel are either paired with the BONE kernel or the LUNG kernel. All images were acquired with a peak kilovoltage of 120-140 kVp, tube current of 40-80 mAs, detector collimation $\leq$ 2.5 mm, pitch of 1.0-2.5, slice thickness of 1.0-3.2 mm and reconstruction interval of 1.0-2.5 mm[33,34]. Although subjects have scans reconstructed using the same projection data, the spatial alignment captured in the field-of-view (FOV) might be different. Therefore, while selecting paired reconstruction kernels, we conducted a manual quality assurance to include scans that had the same spatial alignment [25]. For every reconstruction kernel, we choose 100 volumetric scans that were sampled at random without replacement to train our model. A detailed list of the reconstruction kernels and the number of images used for training have been provided in **Table 1**. We select 100 volumes few or validation to select the optimal checkpoint for generalization. We perform inference on 240 volumes from the withheld test dataset for every reconstruction kernel based on performance on the validation dataset.

## 2.2 Preprocessing

We convert our CT scans from DICOM to NIfTI using the dcm2niix tool (version 1.0.2)[35]. The models were trained on axial slices of size 512 × 512 pixels. In our previous work[30], we observed that the display field-of-view (DFOV) for the Siemens and GE scans were different, resulting in artefacts that were generated outside the DFOV in the harmonized GE scans. Therefore, we apply a circular mask of size 512 × 512 pixels to the Siemens and Philips scans, standardizing the DFOV to be circular for all kernels. We clip the masked images in the range of [-1024, 3072] Hounsfield units (HU) and normalize them to the range of [-1,1][25].

**2.3 Model architecture: Multipath cycle GAN**

**Figure 2** illustrates the two-stage approach for the multipath cycleGAN. In Stage one of training, we use images from four different kernels obtained from the Siemens and GE vendors. We train the encoders, decoders and discriminators across all possible combinations of reconstruction kernels. In Stage two of training, we introduce images reconstructed from three different kernels obtained from the Philips and GE vendors. The new encoders, decoders and all domain specific discriminators are trained while the Stage one encoders and decoders are frozen to enforce supervised domain adaptation. We train across all possible combinations that are made between the Stage one and Stage two kernels.

We build our multipath cycleGAN model based on the traditional cycleGAN[22] model. In a cycleGAN, there exists a forward and backward path to perform image-to-image translation between two domains using two generators and discriminators: one for each direction. However, when new domains are introduced, multiple cycleGAN models are required to be trained between every source and reference. We enable multiple cycleGAN models to be trained across all possible combinations of reconstruction kernels using a multipath cycleGAN that comprises of domain-specific encoders, decoders and discriminators. We build each domain specific generator by breaking down a ResNet model into its corresponding source encoder and target decoder pair that performs image translation through a shared latent space. The ResNet model is a fully convolutional model, adopted from the traditional cycleGAN model, which itself is based on the architecture proposed by Johnson et al.[36]. The ResNet model uses strided and fractionally strided convolutions to upsample and downsample the image. The first layer of the encoder uses a 7×7 Convolution-InstanceNorm-ReLU layer with a stride of one followed by two 3×3 Convolution-InstanceNorm-ReLU layers with a stride of two. The shared latent space is composed of nine residual blocks where each residual block is a convolution block with skip connections comprising of two 3×3 convolution layers, instance normalization and ReLU layers. The decoder uses two 3×3 fractional strided

Convolution-InstanceNorm-ReLU layers followed by a 7×7 fractional strided Convolution-InstanceNorm-ReLU layer. We include additional details of the ResNet model in **Supplementary figure S1**. Every domain specific source encoder maps the input to a shared latent space, resulting in a feature vector of size N × 256 × 128 × 128 where N denotes the batch size. The domain specific target decoder translates the latent feature vector to an image in the style of the target domain The discriminator is a PatchGAN[21] that classifies 70 × 70 patches of images as real or synthetic. The implemented discriminator model is a fully convolutional network consisting of three blocks where each block consists of 4×4 Convolution-InstanceNorm-LeakyRELU layers with a stride of two. A convolution is applied after the last layer to obtain a single channel prediction map. The 70 ×70 patches are generated because of the implicit receptive field that arises from the convolution operation. The setup described facilitates encoder-decoder architectures that stitch all possible reconstruction kernel pairs for multi domain image-to-image translation.

For any given combination of reconstruction kernels, we construct a cycleGAN using the corresponding source and target encoders, decoders and discriminators. We illustrate one such path that involves images reconstructed with the BONE and STANDARD kernels **in Figure 3**. In the forward path, the BONE encoder maps the real BONE image into the shared latent space and the STANDARD decoder produces a synthetic STANDARD image while in the backward path, the STANDARD encoder maps the real STANDARD image through the shared latent space and the BONE decoder produces a synthetic BONE image. The BONE and STANDARD discriminators evaluate whether the images produced in the cyclic path are real or synthetic. We introduce an "identity" loss function, denoted by $\mathcal{L}_{identity}$ to the cycleGAN objective to prevent a shift in the radio-opacity of the synthetic image. This loss function is computed as the mean squared error between the down sampled versions of the real and synthetic images in the forward and backward paths. Down sampling the images smoothens the kernel, thereby allowing the loss to guide the model in preserving the radio-opacity.

## 2.4. Two stage training approach

In Stage one, we train our models using images reconstructed using the Siemens B50f, Siemens B30f, GE BONE and GE STANDARD kernels. The encoder-decoder pairs are trained across 12 different paths for 6 possible pairs of reconstruction kernels. We train this model in parallel on two NVIDIA RTX A6000 GPUs for 120 epochs using a batch size of 2 and an Adam[37] optimizer for the generator and discriminator. The learning rate was 0.002 for the first 60 epochs and decayed linearly to zero for the remaining 60 epochs. We continue training our model from epoch 120 to epoch 200 with a fixed learning rate of 0.002 to obtain additional checkpoints for evaluation. The generator and discriminator are governed by an adversarial loss, implemented using the LSGAN[38] loss function. A cycle consistency loss is implemented to ensure domain consistency during image translation. We weight the cycle consistency loss $\lambda_{cycle}$ to 10 for the forward and backward path, following the default cycleGAN. A squared L2 norm identity loss is computed between the down sampled versions of the real and synthetic images. The real and synthetic images are down sampled to 256 × 256 pixels using bilinear interpolation. This identity loss is weighted by a factor $\lambda_{identity}$. In the first epoch, $\lambda_{identity}$ is set to $10^6$ such that the generators behave as identity networks, producing a reconstructed version of the input. In the subsequent epochs, we scale down the weighting factor by 100 and fix $\lambda_{identity}$ to 0.01 at the sixth epoch such that the adversarial loss takes over the training. We implement a total of 12 adversarial losses, 12 cycle losses and 12 identity losses in Stage one.

In Stage two, we introduce images reconstructed using the GE LUNG, Philips D and Philips C kernels. The GE STANDARD kernels paired with the GE LUNG kernels are used in Stage two of model training. We freeze the encoders and decoders for the Siemens B50f, Siemens B30f, GE BONE and GE STANDARD kernels. The encoders and decoders from Stage one are frozen to promote domain adaptation of the new kernels to the pre trained kernels. We load the trained weights for the encoders, decoders and discriminators trained in Stage one based on the performance of the best epoch on the validation dataset. The new models introduced in Stage two are initialized with the average of the pre-trained weights using the respective encoders, decoders and discriminators. The encoder-decoder pairs are trained across 30 different paths for 15 possible pairs of reconstruction kernels. With the presence of many images for every domain, training across all datasets was prohibitively expensive. Therefore, we randomly sampled 20% of the images from every available dataset in every epoch and trained using the same hardware and optimization settings as Stage one. The learning rate was 0.002 for the first 100 epochs and decayed linearly to zero for the remaining 100 epochs. We implemented 30 adversarial losses, 30 cycle losses and 30 identity losses in Stage two.

**2.5 Baseline models**

We implement the traditional cycleGAN[22] model and the switchable cycleGAN model[26] as baselines for comparison against our model. We implement four individual cycleGAN models for the paired kernels and five individual cycleGAN models for the unpaired kernels that correspond to the path that harmonizes to a reference Siemens B30f kernel for every baseline. For the paths that involve harmonization of soft kernels to a reference hard kernel, we train three individual cycleGAN models and switchable cycleGAN models. We trained each traditional cycleGAN on a single NVIDIA RTX A6000 GPU for 120 epochs using a batch size of six, Adam optimizer for the generator and discriminator and learning rate of 0.002. We use a linear learning rate scheduler to tune the optimizer where the learning rate remains constant for the first 60 epochs and decays linearly till it reaches zero for the remaining 60 epochs. With the availability of minimal checkpoints for the baselines, we continue training the cycleGAN model from epoch 120 to epoch 200 with a fixed learning rate of 0.002 to obtain additional checkpoints for model evaluation. We trained every switchable cycleGAN model on an NVIDIA Quadro RTX 5000 GPU for 200 epochs using the default model configurations which involved a batch size of 16, patch size of 128 × 128 pixels, Adam optimizer for the generator and discriminator and the learning rate of $1 \times 10^{-5}$. The models trained for soft to hard kernel harmonization followed the same configuration.

**3. Performance and statistics**

**3.1 Emphysema quantification**

We assess the efficacy of harmonization for the baselines and multipath cycleGAN model using percent emphysema quantification. We automatically identify regions in the lungs and obtain lung masks using a publicly available algorithm[39]. The density of areas affected with emphysema on CT images ranges from -900 to -1024 HU[40]. We quantify the percentage of voxels in the lung region that have a radio-opacity less than –950 HU using the available lung masks and obtain percent emphysema scores. This approach of thresholding to compute percent emphysema is also known as low attenuation area (LAA). A threshold of –950 HU is considered optimal since it reflects the macroscopic pathological features of emphysema[41,42]. For the paired kernels, we harmonize to the corresponding paired soft kernel while for unpaired kernels, we harmonize from a source kernel to a reference Siemens B30f target kernel and measure percent emphysema.

**3.2 Epoch selection for model selection**

To choose an epoch for model generalizability, we use emphysema quantification as a validation metric for all models. In Stage one, we consider Siemens B50f to Siemens B30f (paired), GE BONE to Siemens B30f (unpaired) and GE STANDARD to Siemens B30f (unpaired), for the purpose of epoch selection. We compute mean squared error (MSE) between the emphysema scores obtained on the paired kernels and Kullback Leibler (KL) Divergence between the emphysema scores obtained on the unpaired kernels. When performing model selection on validation data, we considered both point-to-point matching and distribution matching as our criteria, placing greater emphasis on measurement with paired data with larger coefficients. This design choice ensures that the selected model balances the ability to mitigate differences in emphysema measurements for both paired and unpaired kernels. We rank the MSE and KL divergence scores using the *rankdata* function from the *scipy* python package and perform a weighted sum of ranks using the formula:

$$Epoch_{stage1} = 0.5 * MSE_{B50ftoB30f} + 0.25 * KL_{BONEtoB30f} + 0.25 * KL_{STANDARDtoB30f} \quad (1)$$

In Stage two, we consider Philips C to Siemens B30f (unpaired), Philips D to Siemens B30f (unpaired), GE LUNG to B30f (unpaired), GE LUNG to GE STANDARD (paired) and Philips D to Philips C (paired). The coefficients for MSE and KL are balanced to ensure equal contribution to weight selection. We rank the MSE and KL divergences using the formula:

$$Epoch_{stage2} = 0.2 * MSE_{LUNGtoSTANDARD} + 0.2 * MSE_{DtoC} + 0.2 * KL_{DtoB30f} + 0.2 * KL_{CtoB30f} + 0.2 * KL_{LUNGtoB30f} \quad (2)$$

In the case of soft to hard kernel harmonization, for Stage one, we choose Siemens B30f to B50f (paired) and STANDARD to B50f (unpaired) while for Stage two we choose Philips C to Siemens B50f (unpaired) using the following equations:

$$Epoch_{stage1} = 0.5 * MSE_{B30ftoB50f} + 0.5 * KL_{STANDARDtoB30f} \qquad (3)$$

$$Epoch_{stage2} = KL_{CtoB50f} \qquad (4)$$

Using the above equations, we obtain the overall combined ranks and select the epoch with the lowest rank as the model with the best performance.

### 3.3 Statistical methods

We quantify emphysema measurements using the method described in **Section 3.1** and obtain percent emphysema scores on the hard kernel image, soft kernel image and the harmonized image. We observe the agreement between percent emphysema measurements before and after harmonization on all the paired reconstruction kernel images using Bland-Altman[43] style plots. We compute the median root mean square error (RMSE) and 95% confidence intervals (CI) using bootstrapping[44]. We use 1000 resamples where a random sample is obtained with replacement to form the bootstrap distribution. For every resample, the test statistic was computed, and the confidence intervals of the bootstrap distribution was obtained. For the unpaired reconstruction kernels, we assess statistical significance between emphysema measurements before and after harmonization using the Mann-Whitney U test[45]. After harmonizing all source kernels to a reference target soft kernel, we investigate the impact of age, sex, current smoking status and kernel on the emphysema measurement using a general linear model given by the equation:

$$Y \sim \beta_0 + \beta_1 * X_1 + \beta_2 * X_2 + \beta_3 * X_3 + \beta_4 * X_4 + \varepsilon \qquad (3)$$

where Y is the percent emphysema measurement, $\beta_0$ is the intercept, $X_1, X_2, X_3, X_4$ represent age, sex, reconstruction kernel and current smoking status of the subjects, $\beta_1, \beta_2, \beta_3, \beta_4$ are regression coefficients and $\varepsilon$ denotes error.

## 3.4 Assessment of anatomical consistency before and after harmonization using TotalSegmentator

When training a cycle GAN model to translate an image from one domain to another, the use of distribution matching losses can lead to hallucinations where the model can introduce or remove features[46]. Therefore, it is necessary to quantify hallucinations that occur during harmonization of unpaired reconstruction kernels due to anatomical variations across subjects scanned on two different vendors. We assess consistency of the lung vessels, skeletal muscle and subcutaneous adipose tissue (SAT) before and after harmonization using TotalSegmentator[29]. Specifically, we use the tissue-types model of TotalSegmentator and the lung vessels model[47], generating segmentations of lung vessels, muscle and SAT for the baselines and multipath cycleGAN across all the paths that harmonize a source kernel to a reference Siemens B30f kernel. We compute the Dice scores for lung vessels, muscle and SAT between the segmentations obtained from the same image before and after harmonization. We quantify the effect of the hallucinations between the baselines and multipath cycleGAN model by computing the Cohen's d statistic[48] between the respective lung muscles, skeletal muscle and SAT Dice scores.

## 4. Results

We investigate the hypotheses mentioned in Section 1 by comparing our proposed multipath cycleGAN model with a shared latent space to a standard cycleGAN and a switchable cycleGAN model that use domain specific-latent spaces. Our evaluation focuses on paired and unpaired harmonization scenarios using downstream emphysema quantification. Additionally, we quantitatively evaluated the extent of anatomical hallucinations of different methods. Using the selection criteria described in **Section 2.7**, we selected the 100$^{th}$ epoch and the 119$^{th}$ epoch for Stage one and Stage two of the multipath cycleGAN model. For the standard cycle GAN model, we selected the 120$^{th}$ epoch and 186$^{th}$ epoch while for the switchable cycleGAN model, we selected the 190$^{th}$ epoch and 90$^{th}$ epoch for Stage one and Stage two kernels respectively.

In the case of harmonization from soft to hard kernels, we selected the 80$^{th}$ and 100$^{th}$ epoch from Stage one and two for the multipath cycleGAN. For the standard cycleGAN model, we selected the 162$^{nd}$ and 118$^{th}$ epoch, and the 70$^{th}$ and 145$^{th}$ epoch for the switchable cycleGAN model.

**4.1 Harmonization of paired reconstruction kernels**

We assess the impact of kernel harmonization on percent emphysema quantification for the kernels obtained from the Siemens, GE and Philips vendors. We measure agreements between emphysema measurements before and after harmonization with the help of the RMSE and 95% confidence intervals.  We present all RMSE values and 95% confidence intervals for the paired data are presented in **Table 2**.

Across all pairs, harmonization reduced the median RMSE substantially. The largest improvement in RMSE was seen in the (LUNG, STANDARD) pair across all models with the cycleGAN achieving the largest reduction in RMSE of 96.71%. On the (BONE, STANDARD) pair, the multipath cycleGAN obtained the largest reduction in RMSE (91.52%) while the cycleGAN achieved the largest reduction in RMSE on the (B50f, B30f) and (C, D) pairs. We visualize the RMSE measurements using a Bland Altman style plot for the (B50f, B30f) pair in **Figure 4** with the corresponding plots for the remaining pairs in the supplementary material. Wilcoxon signed rank tests conducted between the measurements before and after measurements were statistically significant ($p<0.05$).

**4.2 Harmonization of all unpaired reconstruction kernels to a reference soft kernel**

We harmonize all the source kernels to the reference Siemens B30f soft kernel. While the Siemens B50f kernel has paired data with the reference, all other source kernels are unpaired. We visualize the emphysema distributions before and after harmonization using boxplots overlaid with strip plots as seen in **Figure 5**. We represent all measurements as the median emphysema measurements. The reference Siemens B30f kernel, had a median of 6.60%. Before harmonization, Siemens B50f, GE BONE, GE LUNG and Philips D overestimated emphysema measurements with median scores of 20.02%, 12.75%, 21.01% and 18.87%. In contrast, GE STANDARD and Philips C had measurements of 2.34% and 2.96%, underestimating emphysema compared to the reference.

After harmonization, all models reduced variability in emphysema measurements and brought most distributions closer to the reference. The cycleGAN aligned the emphysema distributions for Siemens B50f, GE BONE, GE STANDARD, Philips D and Philips C close to the reference median but was less effective for GE LUNG (median 2.84%). The switchable cycleGAN showed improvements for Siemens B50f, GE STANDARD and Philips C but was less effective on GE BONE (median 10.01%), GE LUNG (median 3.28%) and Philips D (median 5.74%). The multipath cycleGAN model showed improvements across Siemens B50f, GE BONE, GE STANDARD and GE LUNG but overcorrected Philips D (median 10.13%) and Philips C (9.13%).

**4.3 Harmonization of all soft kernels to a reference hard kernel**

We harmonize all soft kernels to the reference Siemens B50f hard kernel where the Siemens B30f kernel was paired with the reference. The median emphysema scores for the reference B50f kernel was 20.02%. Before harmonization, the median scores for Siemens B30f, Philips C and GE STANDARD were 6.60%, 2.96% and 2.34% respectively (supplementary **Figure S8**). After harmonization, the distribution of all the soft kernels was closer to the reference hard kernel. The cycleGAN and multipath cycleGAN showed similar trends in performance on all the kernels with minimal differences in the median emphysema score. However, the switchable cycleGAN model underperformed by a small margin on the Siemens B30f and GE STANDARD kernels but achieved a median closer to the reference for the Philips C kernel (median 18.10%) when compared to the other models.

**4.4 Qualitative results**

In the case of paired and reconstruction kernels, we consider a specific region of the lung parenchyma for a subject having scans reconstructed with a hard and soft kernel within a given vendor. Harmonization from a hard kernel to a soft kernel within a given vendor enforces consistency in the texture of the selected lung region. The baseline approach requires models bespoke to each pair. In contrast, our multipath cycleGAN learns a shared latent space enabling harmonization across all paired kernels in the training set as seen in **Figure 6**.

Unpaired reconstruction kernels exhibit entirely different anatomical distributions since two different subjects are scanned on two different scanners. This is reflected in **Figure 7,** depicting anatomical inconsistency and variation in scanner protocols. Our multipath cycleGAN model learns a shared latent space that is capable of harmonizing unpaired kernels to the style of the target soft kernel as seen in the highlighted anatomical region. However, the harmonized GE BONE, GE STANDARD and GE LUNG kernels show subtle changes in the anatomy for the multipath cycleGAN as compared to the cycleGAN and switchable cycleGAN models. Additionally, the cycleGAN baseline model introduces a shift in the anatomy for the harmonized Philips images.

In the case of harmonization from a soft kernel to a hard kernel for unpaired kernels, it is difficult to restore high frequency pixel noise in a soft kernel image because of variability in scanner protocol. Our multipath cycleGAN model and the standard cycleGAN model restored the high frequency details in the highlighted region of interest see in **Figure 8**. The switchable cycleGAN also harmonized the soft kernels to the style of the reference hard kernel. However, on the Siemens kernels, the region of interest is slightly blurry compared to the remaining models.

**4.5 Impact of kernel on emphysema quantification**

We fit a general linear model to observe the effect of reconstruction kernel on percent emphysema quantification while controlling for age, sex and smoking status of the subjects (**Table 3**). Prior to harmonization, hard kernels have positive coefficients indicating higher emphysema scores as compared to the reference Siemens B30f soft kernel while soft kernels have negative coefficients indicating lower emphysema scores as compared to the reference. The Siemens B50f, GE BONE, Philips D and GE LUNG kernels had large positive coefficient while the Philips C and GE STANDARD kernels had negative coefficients. All covariates were significantly different ($p<0.05$) before harmonization. After harmonization, both the baseline and the multipath cycleGAN model mitigated systematic differences in emphysema scores ($p>0.05$) for the Siemens B50f, GE BONE, Philips D and GE LUNG kernels. However, the Philips D kernel continued to remain significant for the multipath cycleGAN after harmonization. While the Philips C and GE LUNG kernels showed an increase in their coefficients after harmonization, they continued to remain significant for baseline models and the multipath cycleGAN. The switchable cycleGAN models eliminated the effects of the Siemens B50f and GE STANDARD kernels. Although the remaining kernels showed an increase or decrease in the coefficients, the effect of kernel on emphysema was partially mitigated and therefore continued to remain significant.

### 4.6 Anatomical consistency after harmonization to a reference soft kernel

We quantify the consistency in anatomy for all unpaired reconstruction kernels between the multipath cycleGAN and the baseline models by computing the effect size between the Dice scores for lung vessels, muscle and SAT using the Cohen's d statistic[48] (**Table 4**). The effect size is small when Cohen's d = 0.2, medium when Cohen's d = 0.5 and large when Cohen's d = 0.8. A positive Cohen's d indicated that group one had a higher mean than group two while a negative Cohen's d indicated the opposite. Cohen's d was greater than 0.8 for all pairs between Dice on muscle and SAT from the multipath cycleGAN and standard cycleGAN model. The range of Dice scores was higher for all pairs for the multipath cycleGAN compared with pairs for the cycleGAN indicating greater anatomical consistency **(Figure 9)**.

When computing the effect size between the Dice scores for muscle and SAT between the multipath cycleGAN and the switchable cycleGAN, the effect sizes were highly variable across all pairs. For the Philips D kernel, the effect size was small. GE BONE showed a small effect size for muscle while SAT showed a large effect. The negative value indicates that the mean dice for the switchable cycleGAN was higher than the multipath cycleGAN. For all the other pairs, there was a medium effect size for muscle and a large effect size for SAT, indicating that the switchable cycleGAN was able to preserve anatomy much better than the cycleGAN.

When compared to the cycleGAN on the lung vessels task, the range of Dice scores for the multipath cycleGAN were higher on the GE BONE, GE STANDARD and GE LUNG kernels compared to the Philips kernels where the effect sizes were small for Philips D and large for Philips C (**Figure 10**). The effect sizes computed on the Dice scores between the multipath cycleGAN and the switchable cycleGAN were high on GE STANDARD, Philips D and Philips C kernels and medium on the GE BONE and GE LUNG images, suggesting that switchable cycleGAN showed better anatomical overlap on most of the kernels.

### 5. Discussion and conclusion

In this work, we investigate CT kernel harmonization of paired and unpaired reconstruction kernels in a low-dose lung cancer screening cohort using a two-stage multipath cycleGAN model that houses a shared latent space. We demonstrate that the shared latent space plays an important role in kernel harmonization that mitigates differences in emphysema measurements. Furthermore, we show that harmonizing all kernels to the reference soft kernel mitigates kernel effects while controlling for age, sex and smoking status. This can be attributed to the ability of the shared latent space to effectively capture meaningful representations across paired and unpaired reconstruction kernels, therefore mitigating variability in emphysema measurements. In the case of unpaired reconstruction kernels, we demonstrated that the shared latent space preserved the overall anatomy outside the lung by comparing the effect size with the baseline models.

Our study investigated the application of a shared latent space in enforcing consistent emphysema measurements across paired reconstruction kernels compared to separate latent spaces used in models like the traditional cycleGAN model. Before harmonization, images reconstructed with hard and soft kernels disagree on emphysema measurements. By harmonizing hard kernels to the style of the soft kernels through a shared latent space, our approach successfully mitigates differences in emphysema measurements, thereby enforcing consistency as seen in **Figure 4** and the supplementary figures. These findings are consistent with prior works such as Bak et al[49]. who demonstrated that LDCT scans reconstructed with the Siemens B50f kernel exhibited thrice the amount of emphysema as compared to images reconstructed using the B31f kernel. Our approach demonstrates superior performance compared to the standard cycleGAN on the (BONE, STANDARD) pair and achieves comparable results on the (B50f, B30f) and (LUNG, STANDARD) pairs. Compared to the switchable cycleGAN, our approach shows better performance on the (B30f, B50f) and (BONE, STANDARD) pairs while maintaining comparable performance on the (LUNG, STANDARD) pair. However, the multipath cycleGAN underperforms on the (C, D) pairs as compared to the baseline models. Overall, these results highlight the ability of the shared latent space to learn a common representation among all paired kernels, therefore enforcing consistent emphysema measurements that are comparable to separate latent spaces.

A key aspect of our study investigates a shared latent space to mitigate kernel effects when all kernels are harmonized to a reference soft kernel. Prior to harmonization, we observe discrepancies across unpaired hard and soft kernels. Our approach shows improvements in the emphysema distribution for most of the kernels when harmonized to the reference Siemens B30f soft kernel as seen in **Figure 5**. The shared latent space was capable of enforcing consistency in measurements across all kernels, thereby allowing for unified multi domain kernel harmonization without the need for paired data. When compared to the baseline models, our approach achieves comparable performances across the kernels employed in Stage one, effectively mitigating the kernel effects on emphysema quantification. For the Stage two training, while the encoders and decoders of the Stage one kernels were frozen, the discriminator, acting as the adversary, facilitated the harmonization of these kernels to the reference soft kernel. While differences are mitigated post harmonization, there is variability in the performance of all models across the GE LUNG, Philips D and Philips C kernels, suggesting that kernel effects were only partially mitigated post harmonization. In the general linear model analysis, coefficients for age, sex and current smoking status showed minor changes, suggesting that the underlying biological meaning of the subject was preserved after harmonization. We also investigated the ability of the shared latent space to mitigate differences in emphysema scores when all soft kernels are harmonized to a reference hard kernel. Our proposed approach could recover the high frequency pixel noise from the soft kernel image across all kernels. In the case of emphysema quantification, we observe comparable performance with the baseline standard cycleGAN model across all the kernels with a slight difference in the scores obtained on the GE STANDARD kernels.

Emphysema quantification is an increasingly utilized method of evaluating CT imaging of the chest and has proven to be an extremely useful tool in the assessment of respiratory pathology. There are several different patterns of emphysema that can be categorized with quantitative analysis. The severity of emphysema can be combined with a quantitative assessment of air trapping which can improve the correlation of CT findings with histologic patterns[50]. Assessing air trapping is important for other disease as well, including interstitial lung disease such as hypersensitivity pneumonitis[51], and is also an essential finding when evaluating patients following lung transplant[52]. However, these quantitative measures are sensitive to the reconstruction kernel. Previous studies have shown that varying the reconstruction algorithm can lead to differences exceeding 15% in percent emphysema, creating a false appearance of disease progression or improvement in longitudinal follow-up[4]. In clinical scenarios such as bronchoscopic lung volume reduction (BLVR), inaccuracies introduced by the kernel can result in patients being mistakenly included or excluded from therapy. Standardization minimizes risks of spurious clinical interpretations, supporting reliable treatment selection and disease monitoring. It is imperative that quantitative assessment continue to improve and be standardized as this is an increasingly important metric that contributes to a growing number of disease processes.

We also examine the ability of the shared latent space to enforce anatomical consistency on all the unpaired kernels after harmonization. CycleGAN models are capable of hallucinations because their objective function involves distribution matching losses that can lead to hallucinations[46]. Using TotalSegmentator, we obtained segmentations of lung vessels, skeletal muscle and subcutaneous adipose tissue and computed the Dice overlap between the non-harmonized and harmonized images. For skeletal muscle and SAT, our results show that the multipath cycleGAN effectively preserved anatomy when compared to the standard cycleGAN model as seen in the Dice score boxplots shown in **Figure 9** and effect sizes presented in **Table 4**. For lung vessels, the multipath cycleGAN showed reasonable overlap before and after harmonization on all the GE kernels as compared to the cycleGAN as observed in **Figure 10.** These results suggest that the standard cycleGAN introduced anatomical shifts in the lung vessels during harmonization while the shared latent space captured consistent anatomy across all paths that highlighted unpaired kernel harmonization. However, we observe that the performance of the multipath cycleGAN falls slightly short in comparison to the switchable cycleGAN. We believe that architecture of the switchable cycleGAN involved a polyphase decomposition UNet with skip connections that helped preserve anatomical consistency, thereby resulting in higher Dice scores.

Our work is not without its limitations. For the images reconstructed with the GE BONE and GE STANDARD kernels, while the overall range of Dice scores is high for muscle and SAT, there exists hallucinations in the images where skeletal muscle around the chest wall and SAT are modified as lung tissues as evidenced in **Figure 8**. Therefore, we did not evaluate body composition measurements for skeletal muscle and SAT. A potential future solution could involve the use of prior anatomical features such as segmentation masks that constrain synthesis and prevent anatomical hallucination. The FOV in CT images can vary based on reconstruction protocol and vendor as illustrated by Xu et al.[53]. We standardized the images to have a circular field of view to enforce consistent field of view across all images. However, an arguably more complete solution that avoids cropping altogether is semantically extending the field of view thereby providing context on muscle and SAT to the model during training. While differences in measurements were mitigated for the Philips D, Philips C and GE LUNG kernels, partial kernel effects continued to remain which requires further exploration.

To summarize our work, we develop a multipath cycleGAN model that uses a shared latent space to harmonize paired and unpaired reconstruction kernels in a low-dose lung cancer screening population with an emphasis on quantitative tasks that include emphysema quantification. Furthermore, we quantify inconsistency in anatomy post harmonization for all unpaired kernels using Dice overlap and Cohen's d effect size. These results demonstrate the promising effectiveness of representing all kernel and vendor pairs in a harmonizing shared latent space. With the advancement in CT technology, the need to train domain specific models is burdensome in multi-centre studies. Our approach provides a unified solution to harmonize paired and unpaired reconstruction kernels that has the potential to advance flexibility, scalability and performance in CT harmonization.

**Acknowledgements**

This research was funded by the National Cancer Institute (NCI) grant R01 CA253923-04, R01 CA 253923-04S1. This work was also supported in part by the Integrated Training in Engineering and Diabetes grant number T32 DK101003. This research is also supported by the following awards: National Science Foundation CAREER 1452485; NCI U01 CA196405; UL1 RR024975-01 of the National Center for Research Resources and UL1 TR000445-06 of the National Center for Advancing Translational Sciences; Martineau Innovation Fund grant through the Vanderbilt-Ingram Cancer Center Thoracic Working Group; NCI Early Detection Research Network grant 2U01CA152662. The Vanderbilt Institute for Clinical and Translational Research (VICTR) is funded by the National Center for Advancing Translation Science Award (NCATS), Clinical Translational Science Award (CTSA) Program, Award Number 5UL1TR002243-03. The content is solely the responsibility of the authors and does not necessarily represent the official views of the NIH. We use generative AI to create code segments based on task descriptions, as well as debug, edit, and autocomplete code. Additionally, generative AI technologies have been employed to assist in structuring sentences and performing grammatical checks. It is imperative to highlight that the conceptualization, ideation, and all prompts provided to the AI originate entirely from the authors' creative and intellectual efforts. We take accountability for the review of all content generated by AI in this work.

and pixel noise in computed tomography. *IEEE Trans Med Imaging*. 2003;22(7):846-853. doi:10.1109/TMI.2003.815073

2. Lasek J, Piórkowski A. CT Scan Transformation from a Sharp to a Soft Reconstruction Kernel Using Filtering Techniques. In: *Communications in Computer and Information Science*. Vol 1376 CCIS. Springer Science and Business Media Deutschland GmbH; 2021:56-65. doi:10.1007/978-981-16-1086-8_6

3. Mackin D, Ger R, Gay S, et al. Matching and Homogenizing Convolution Kernels for Quantitative Studies in Computed Tomography. *Invest Radiol*. 2019;54(5):288-295. doi:10.1097/RLI.0000000000000540

4. Boedeker KL, McNitt-Gray MF, Rogers SR, et al. Emphysema: Effect of reconstruction algorithm on CT imaging measures. *Radiology*. 2004;232(1):295-301. doi:10.1148/radiol.2321030383

5. Gierada DS, Bierhals AJ, Choong CK, et al. Effects of CT Section Thickness and Reconstruction Kernel on Emphysema Quantification. Relationship to the Magnitude of the CT Emphysema Index. *Acad Radiol*. 2010;17(2):146-156. doi:10.1016/j.acra.2009.08.007

6. Troschel AS, Troschel FM, Best TD, et al. Computed Tomography-based Body Composition Analysis and Its Role in Lung Cancer Care. *J Thorac Imaging*. 2020;35(2):91-100. doi:10.1097/RTI.0000000000000428

7. Lessmann N, Van Ginneken B, Zreik M, et al. Automatic Calcium Scoring in Low-Dose Chest CT Using Deep Neural Networks with Dilated Convolutions. *IEEE Trans Med Imaging*. 2018;37(2):615-625. doi:10.1109/TMI.2017.2769839

8. Meyer M, Ronald J, Vernuccio F, et al. Reproducibility of CT radiomic features within the same patient: Influence of radiation dose and CT reconstruction settings. *Radiology*. 2019;293(3):583-591. doi:10.1148/radiol.2019190928

9. Denzler S, Vuong D, Bogowicz M, et al. *Impact of CT Convolution Kernel on Robustness of Radiomic Features for Different Lung Diseases and Tissue Types*.; 2021.

10. Ohkubo M, Wada S, Kayugawa A, Matsumoto T, Murao K. Image filtering as an alternative to the application of a different reconstruction kernel in CT imaging: Feasibility study in lung cancer screening. *Med Phys*. 2011;38(7):3915-3923. doi:10.1118/1.3590363

11. Sotoudeh-Paima S, Samei E, Abadi E. CT-HARMONICA: physics-based CT harmonization for reliable lung density quantification. In: Iftekharuddin KM, Chen W, eds. *Medical Imaging 2023: Computer-Aided Diagnosis*. SPIE; 2023:52. doi:10.1117/12.2654346

12. Christianson O, Winslow J, Frush DP, Samei E. Automated technique to measure noise in clinical CT examinations. *American Journal of Roentgenology*. 2015;205(1):W93-W99. doi:10.2214/AJR.14.13613

13. Zarei M, Sotoudeh-Paima S, McCabe C, Abadi E, Samei E. Harmonizing CT images via physics-based deep neural networks. In: Fahrig R, Sabol JM, Yu L, eds. *Medical Imaging 2023: Physics of Medical Imaging*. SPIE; 2023:66. doi:10.1117/12.2654215
23

# Tables

**Table 1.** Reconstruction kernels from four different manufacturers are used for kernel harmonization. For a given manufacturer, there exists paired reconstruction kernels that have a one-to-one pixel correspondence. However, subjects across manufacturers remained unpaired, resulting in difference in protocol and anatomy. The available data is used to train paired as well unpaired reconstruction kernels in an unpaired manner.

| Manufacturer | Reconstruction kernel | No. of volumes for training | Total number of training slices |
|---|---|---|---|
| Siemens | B50f (hard) | 100 | 16343 |
|  | B30f (soft) | 100 | 16343 |
| GE | BONE (hard) | 100 | 14614 |
|  | STANDARD (soft) | 100 | 14614 |
|  | LUNG (hard) | 100 | 15723 |
|  | STANDARD (soft) | 100 | 15723 |
| Philips | D (hard) | 100 | 19359 |
|  | C (soft) | 100 | 19359 |

**Table 2.** Comparison of difference in percent emphysema measurements between paired reconstruction kernels before and after harmonization using standard cycleGAN, switchable cycleGAN and multipath cycleGAN (ours). We present the root mean squared error (RMSE) computed between the source kernel and the target kernel pair followed by 95% confidence intervals. Kernel harmonization significantly reduces difference in measurements before and after harmonization ($p < 0.05$, Wilcoxon signed-rank test with bootstrapping). Our approach achieves comparable performance with baselines for most of the paired kernels, demonstrating the effectiveness of a shared latent space.

| Pairs | Difference in percent emphysema | | | |
|---|---|---|---|---|
|  | Before harmonization | cycleGAN harmonization | Switchable cycleGAN harmonization | Multipath cycleGAN harmonization (ours) |
| (B50f, B30f) | 12.12 [11.76, 12.42] | 1.16 [1.05, 1.29] | 1.57 [1.41, 1.71] | 1.35 [1.22, 1.52] |
| (BONE, STANDARD) | 9.91 [9.50, 10.33] | 0.91 [0.81, 1.05] | 1.01 [0.89, 1.21] | 0.84 [0.75, 0.95] |
| (D, C) | 15.81 [15.28, 16.33] | 1.31 [1.18, 1.49] | 1.78 [1.57, 2.12] | 2.47 [2.11, 2.93] |
| (LUNG, STANDARD) | 19.20 [18.58, 19.78] | 0.63 [0.52, 0.75] | 0.99 [0.78, 1.33] | 1.05 [0.83, 1.46] |

**Table 3.** Multivariate linear regression model to assess the impact of age, sex, current smoking status and kernel on percent emphysema quantification for unpaired kernel harmonization to a reference Siemens soft kernel. We present regression coefficients, standard error and the *p* value to explain the impact of each variable on emphysema quantification. After harmonization, we observe that the impact of certain kernels on emphysema is mitigated upon harmonization for either the cycle GAN, multipath cycle GAN or both models highlighted in bold. The coefficients for age, sex and current smoking status showed minimal changes and continued to remain post harmonization. Harmonization preserves the biological meaning of the subjects while simultaneously mitigating kernel effects on percent emphysema quantification.

| Covariate | | Coefficient | | | | *p*-value | | |
|---|---|---|---|---|---|---|---|---|
| | Before | After (cycleGAN) | After (Switchable cycleGAN) | After (Multipath cycleGAN) | Before | After (cycleGAN) | After (Switchable cycleGAN) | After (Multipath cycleGAN) |
| Siemens B50f | 11.72 | 0.58 | 0.63 | -0.07 | <0.05 | 0.35 | 0.30 | 0.91 |
| GE BONE | 4.13 | -1.00 | 2.51 | -1.21 | <0.05 | 0.11 | <0.05 | 0.06 |
| Philips C | -4.00 | 1.56 | -1.70 | 2.20 | <0.05 | 0.01 | 0.01 | <0.05 |
| Philips D | 11.14 | 0.19 | -4.50 | 2.94 | <0.05 | 0.77 | <0.05 | <0.05 |
| GE LUNG | 12.78 | -4.44 | -3.93 | -2.26 | <0.05 | <0.05 | <0.05 | <0.05 |
| GE STANDARD | -5.09 | -1.14 | -0.40 | -0.70 | <0.05 | 0.07 | 0.51 | 0.28 |
| Age | 0.20 | 0.18 | 0.20 | 0.22 | <0.05 | <0.05 | <0.05 | <0.05 |
| Sex | -1.76 | -1.83 | -2.08 | -2.05 | <0.05 | <0.05 | <0.05 | <0.05 |
| Current smoking status | -3.18 | -2.30 | -2.42 | -2.68 | <0.05 | <0.05 | <0.05 | <0.05 |

**Table 4.** Distribution matching losses in cycleGAN models introduce hallucinations. We quantify hallucinations by measuring the effect size between Dice scores obtained on segmentations of skeletal muscle and subcutaneous adipose tissue (SAT) before and after harmonization using the Cohen's d statistic. Furthermore, we measure the effect size between dice scores obtained on segmentations of lung vessels. Cohen's d quantifies the magnitude of difference between the Dice scores that signifies the degree of hallucinations induced post harmonization. We present the Cohen's d statistic for all unpaired kernels that are harmonized to a reference Siemens B30f soft kernel. The effect size is small when Cohen's d = 0.2, medium when Cohen's d = 0.5 and large when Cohen's d = 0.8. A positive Cohen's d indicated that group one had a higher mean than group two while a negative Cohen's d implied the opposite. Our approach maintains anatomical consistency as compared to the cycleGAN model. However, the switchable cycleGAN model does better on select kernels compared to our approach. For the lung vessels, our approach shows reasonable overlap on most of the kernels compared to the cycleGAN, while the switchable cycleGAN shows good overlap compared to out proposed approach.

| Harmonized kernel | Cohen's d for skeletal muscle | | Cohen's d for SAT | | Cohen's d for lung vessels | |
|---|---|---|---|---|---|---|
| | Multipath cycleGAN (ours) vs cycleGAN | Multipath cycleGAN (ours) vs switchable cycleGAN | Multipath cycleGAN (ours) vs cycleGAN | Multipath cycleGAN (ours) vs switchable cycleGAN | Multipath cycleGAN (ours) vs cycleGAN | Multipath cycleGAN (ours) vs switchable cycleGAN |
| GE BONE -> Siemens B30f | 1.54 | -0.15 | 0.81 | -1.08 | 6.42 | 0.77 |
| GE STANDARD -> Siemens B30f | 3.99 | -1.35 | 1.35 | -1.61 | 24.90 | -5.48 |
| Philips D -> Siemens B30f | 3.12 | 0.39 | 2.58 | 0.20 | -0.14 | -0.94 |
| Philips C -> Siemens B30f | 3.34 | -1.56 | 2.54 | -1.55 | -1.05 | -4.81 |
| GE LUNG -> Siemens B30f | 2.30 | -0.39 | 1.60 | -0.66 | 7.49 | -0.42 |

**Figures**

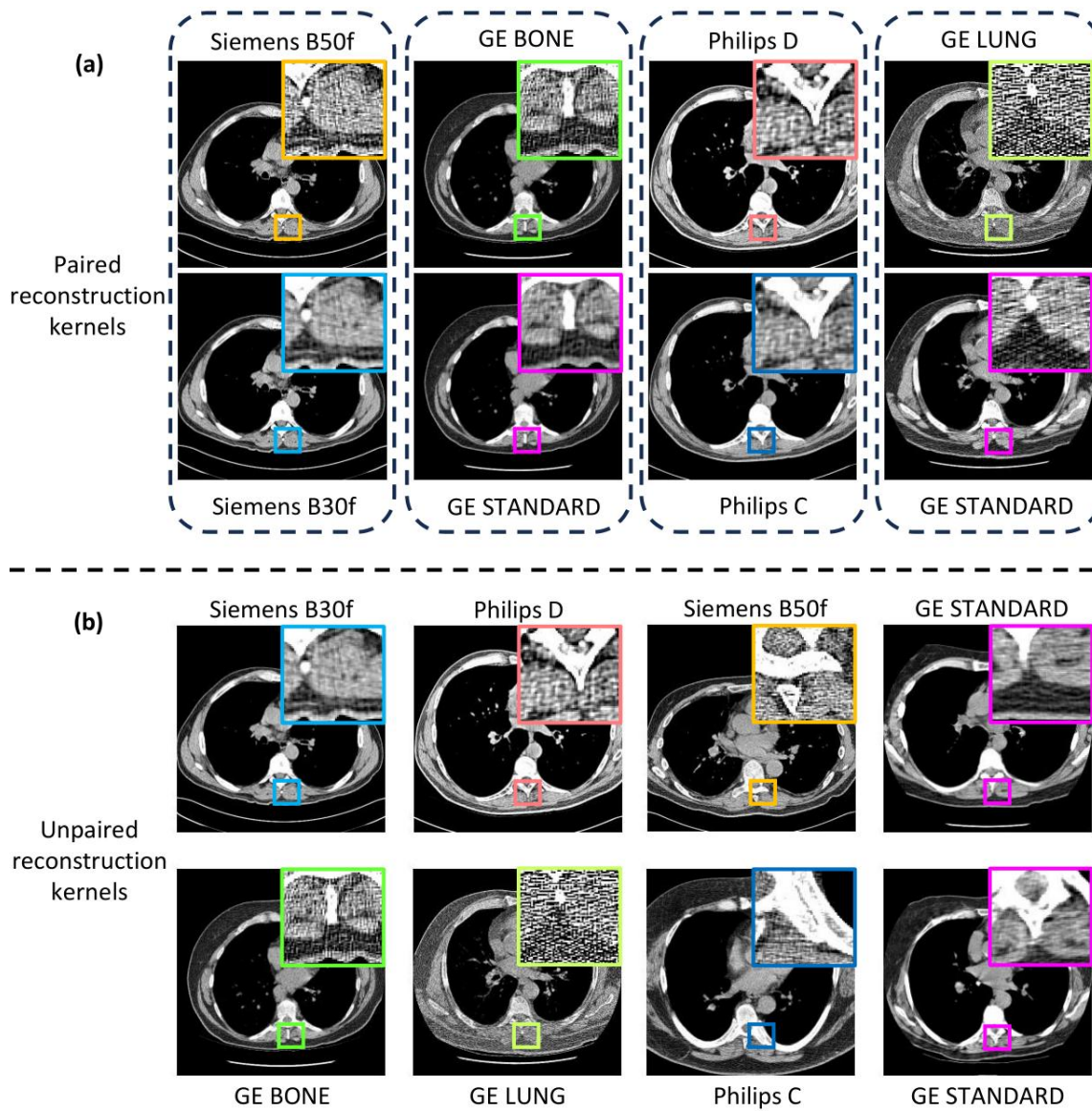

**Figure 1:** Reconstruction kernels influence the noise and resolution of the underlying anatomical structure in a computed tomography image. **(a)** Paired reconstruction kernels obtained from a given vendor exhibit a one-to-one pixel correspondence between the scans which enables kernel harmonization. However, **(b)** across vendors, unpaired kernels show differences in anatomy, scan protocol, field of view and reconstruction window. This creates additional difficulties that make harmonization a more challenging task.

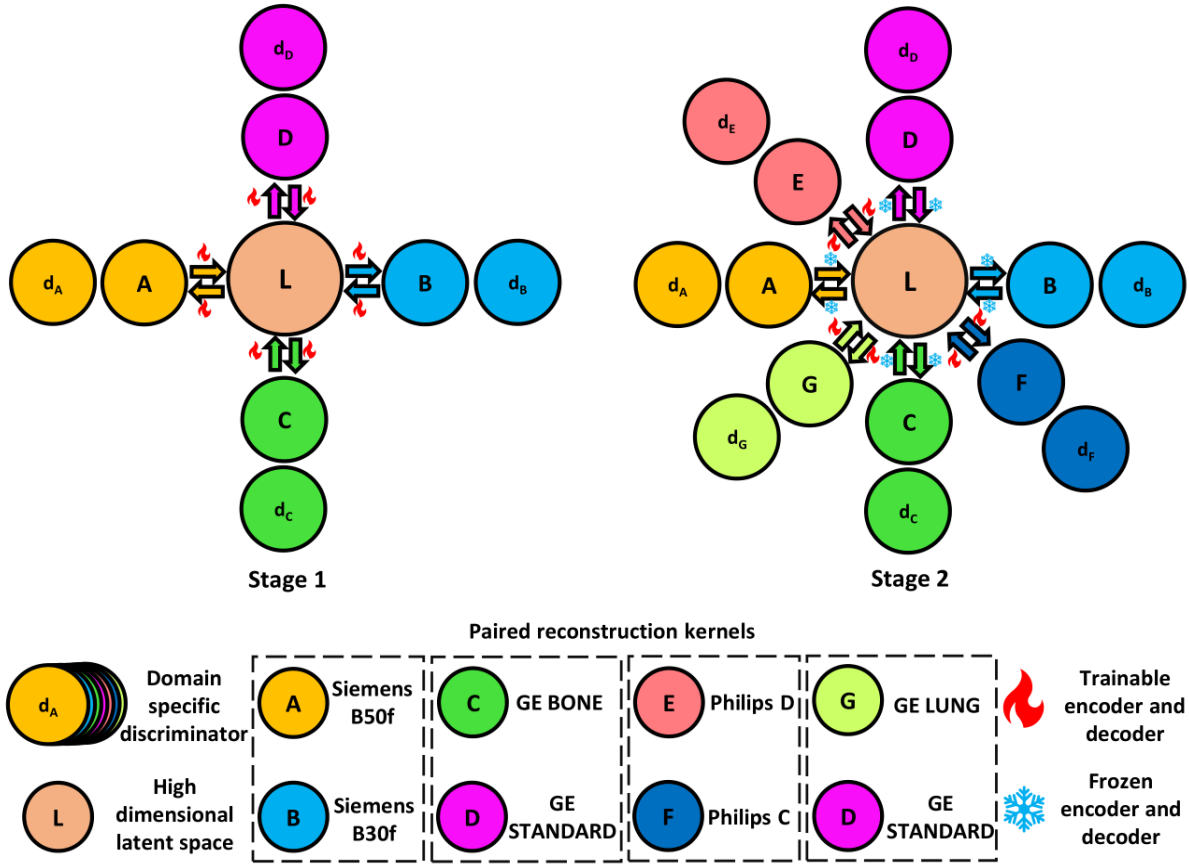

**Figure 2:** We hypothesize that inter-vendor and intra-vendor harmonization across different combinations of reconstruction kernels can be performed using a multipath cycleGAN model, trained in two stages. The paired reconstruction kernels are grouped together based on the manufacturer while all other combinations denote unpaired kernels. In Stage one, we harmonize across all possible combinations of four different reconstruction kernels through a high dimensional shared latent space (denoted by L) using domain specific encoder-decoder architectures. In Stage two, we freeze the trained encoders and decoders and train the model to account for new combinations across all available reconstruction kernels.

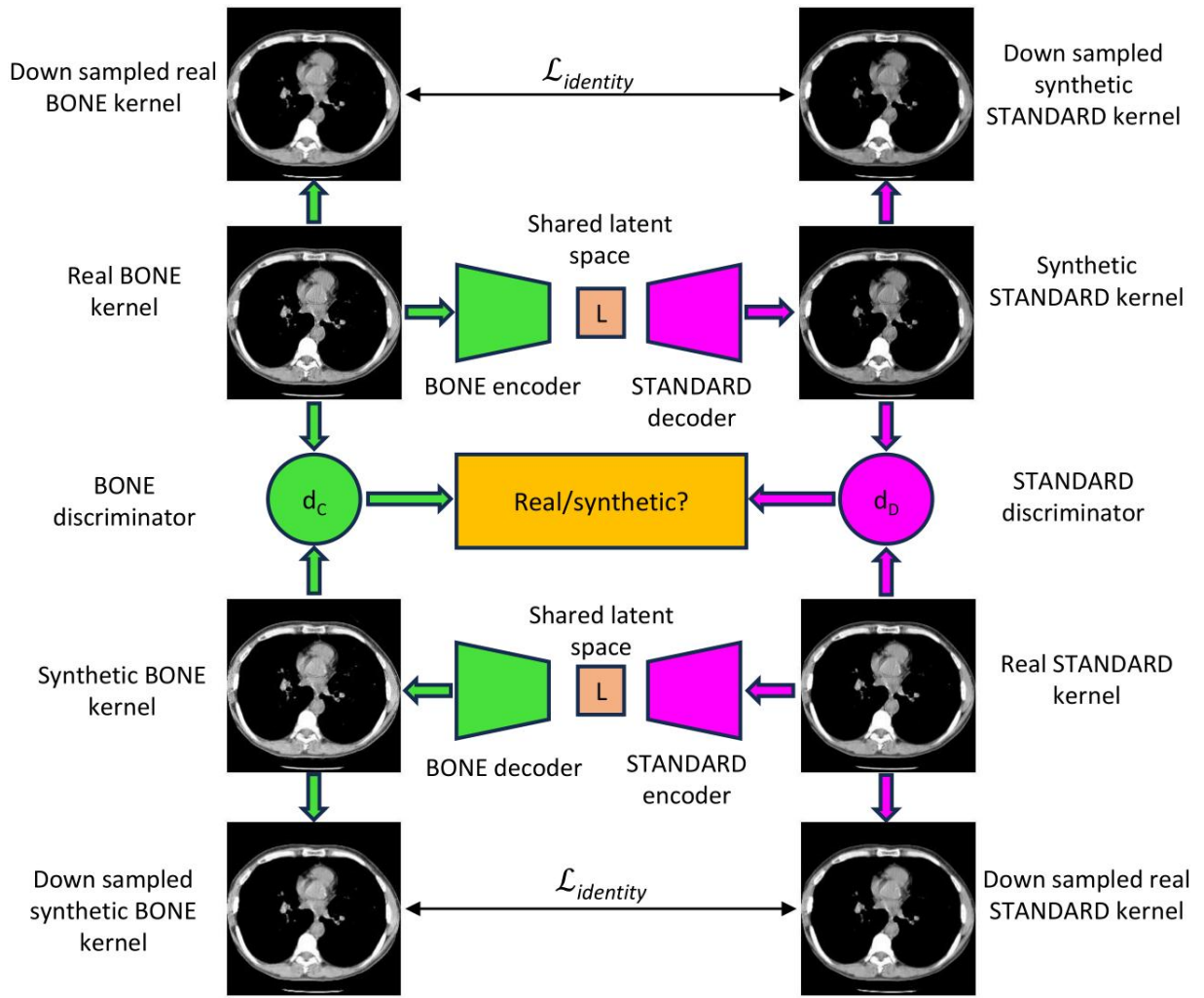

**Figure 3:** For any given pair of reconstruction kernels, there exists a forward and backward path. The generator is a ResNet, formed by a source encoder and target decoder in the forward path and a target encoder and source decoder in the backward path. Each generator produces a synthetic image with the style of either the source or target kernel. A PatchGAN is used as a discriminator for the corresponding domain to distinguish between real and synthetic images. In addition to the adversarial and cycle consistency losses, an identity L2 loss is applied between the down sampled real and synthetic image to prevent a shift in the radio-opacity of the images.

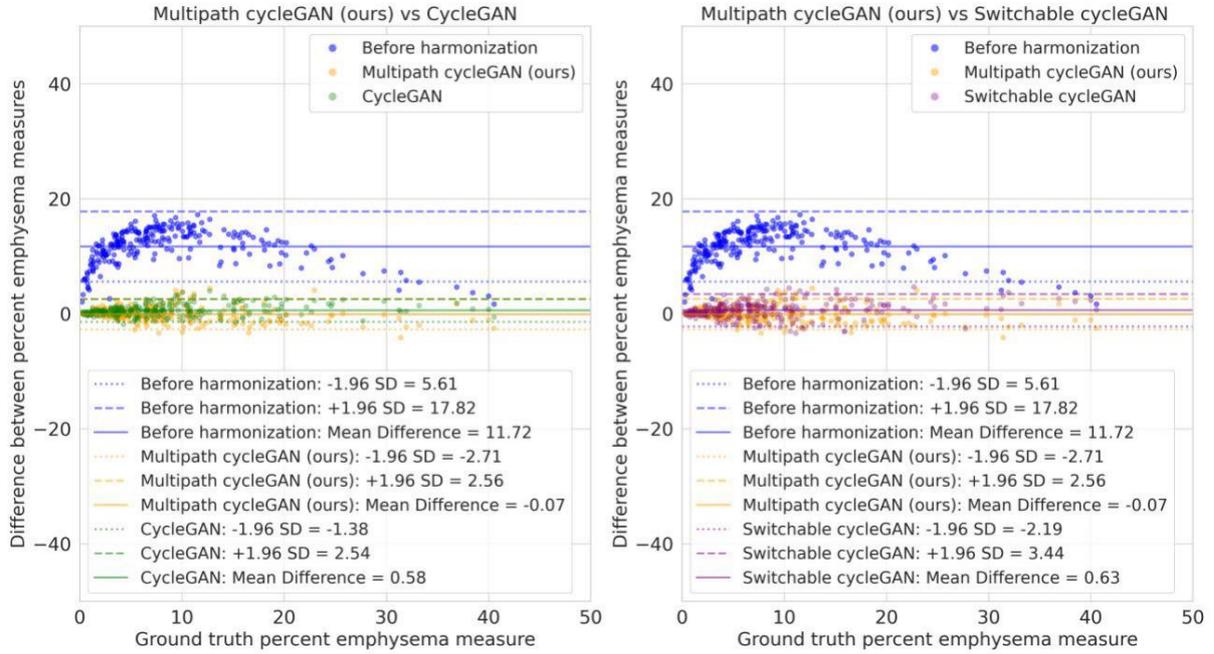

**Figure 4.** Bland Altman style plots are used to represent the agreement on emphysema measurements between paired hard and soft reconstruction kernels before and after harmonization. The solid line represents the mean difference between the emphysema measurements and the dashed lines represent the standard deviation (SD) as 95% confidence intervals [-1.96SD, +1.96SD]. Blue represents measurements without harmonization, yellow represents the multipath cycleGAN, green represents the standard cycleGAN and purple represents the switchable cycleGAN. Emphysema measurements disagree before harmonization. After harmonization, all models show improvements in agreements. The mean difference is close to zero, with the multipath cycleGAN showing better agreement on the Siemens kernels.

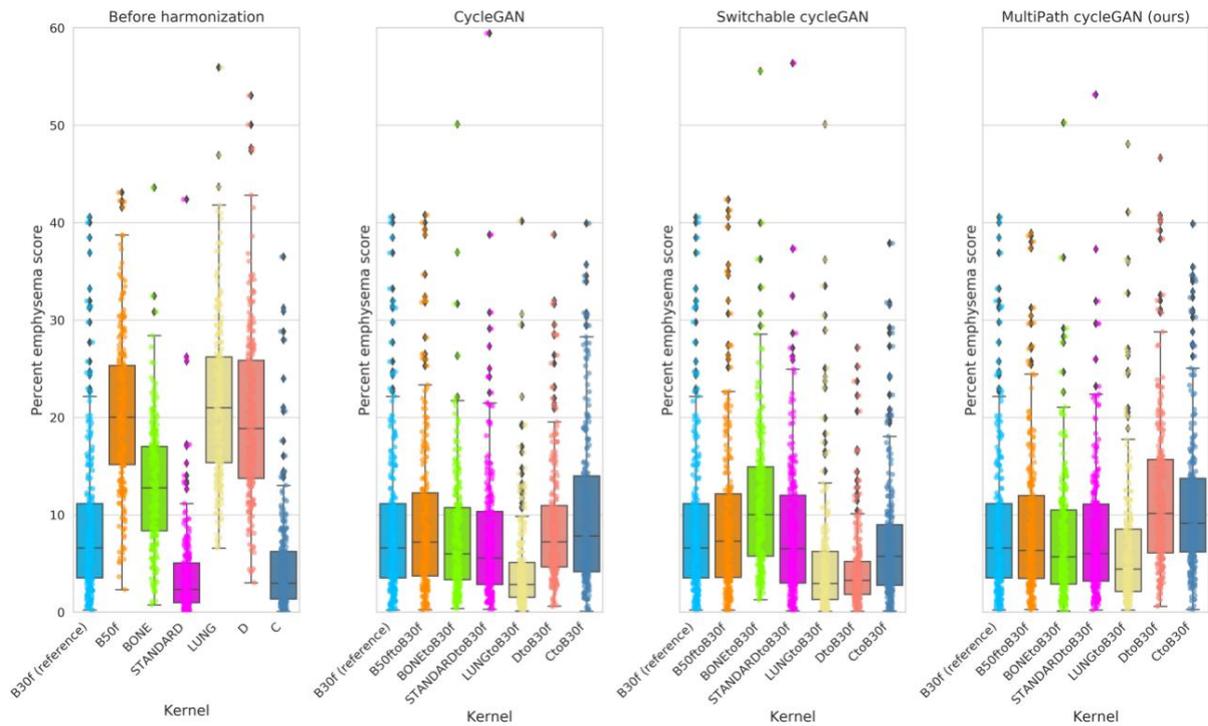

**Figure 5.** Unpaired CT kernels are challenging to harmonize due to the difference in scanner protocol, anatomy and field of view. Before harmonization, hard kernels have a higher range of emphysema scores compared to soft kernels that are preferred for emphysema. The reference Siemens B30f kernel is represented by the light blue box and strip plot. Harmonization of all kernels to the reference kernel mitigates differences in emphysema measurement as evidence by closer medians to the reference. The cycleGAN models show the best performance in mitigating differences in measurements across kernels, followed by the multipath cycleGAN and switchable cycleGAN models.

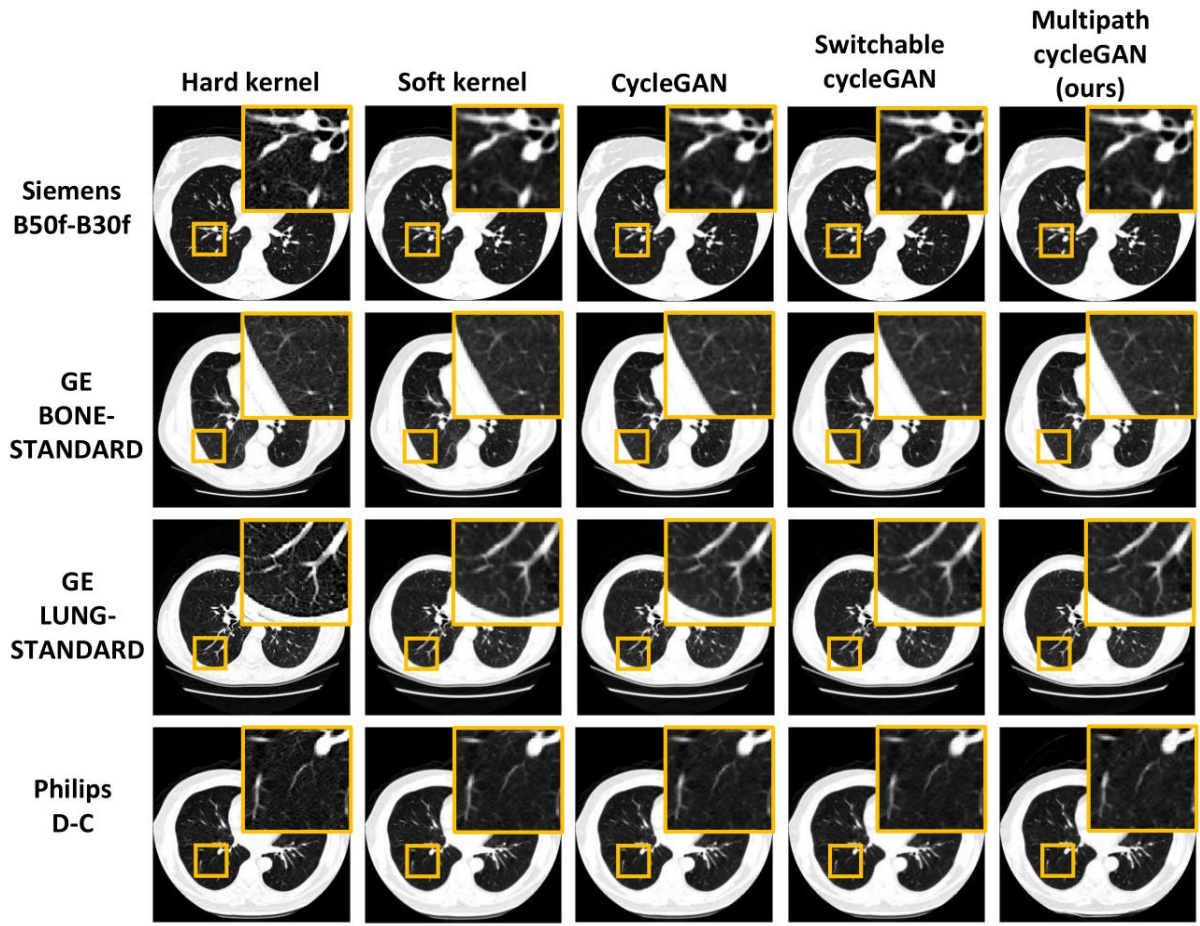

**Figure 6.** Paired kernels exhibit a one-to-one pixel correspondence between the hard and soft kernel in each vendor with differences in the pixel noise. Hard kernels sharpen the image while soft kernels smoothen the image. Harmonizing to the corresponding soft kernel enforces qualitative consistency in the lung parenchyma for the multipath cycleGAN that is comparable to the baseline models.

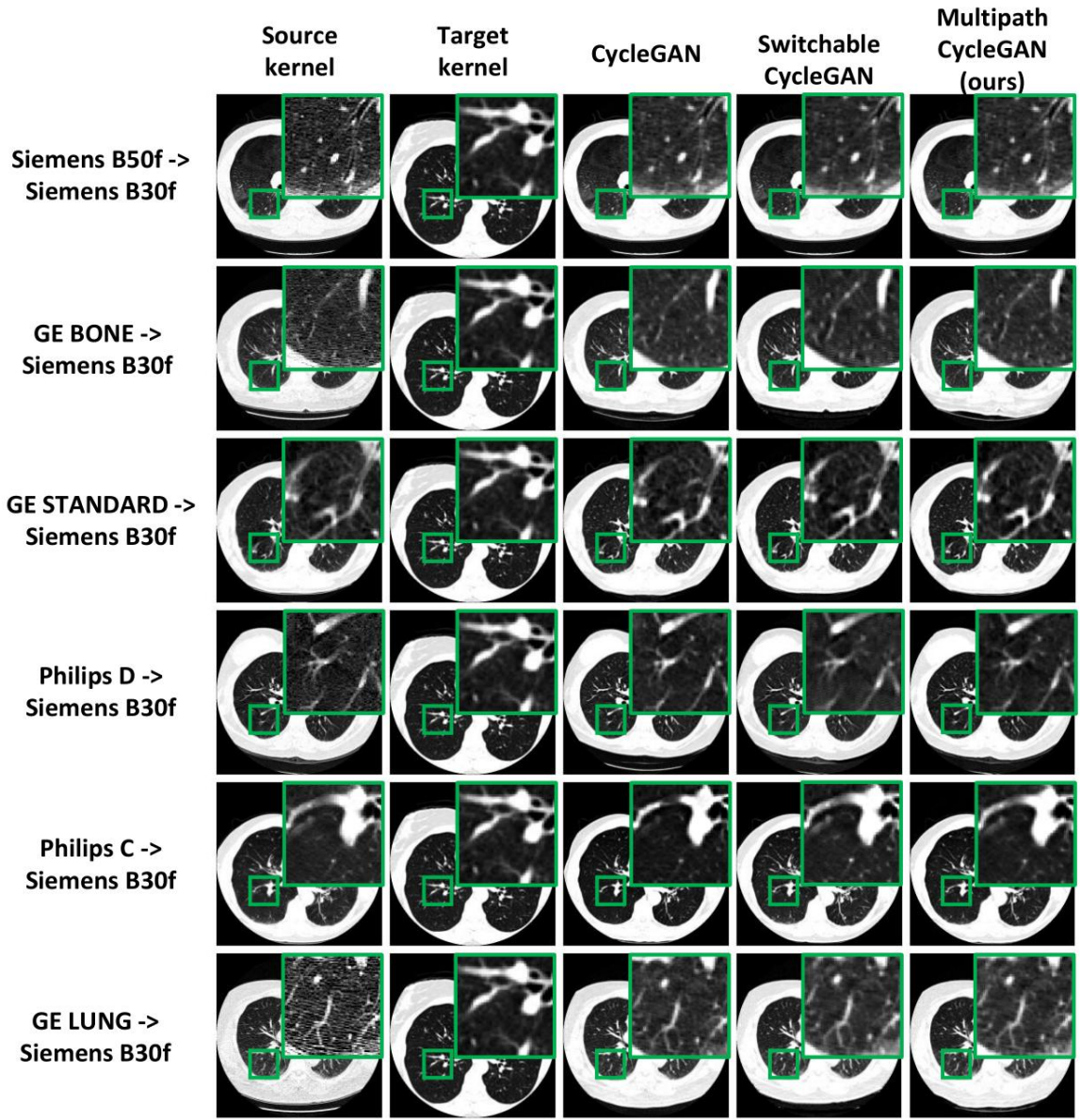

**Figure 7.** Variation in scanner protocol introduces differences in the texture of the lung parenchyma, thereby introducing differences in quantitative image measures. Harmonization of all kernels to the reference B30f soft kernel enforces consistency in the texture of the lung parenchyma which benefits downstream tasks that include percent emphysema quantification. Multipath cycleGAN enforces consistency in texture of the lung parenchyma, comparable to baseline models across all unpaired kernels.

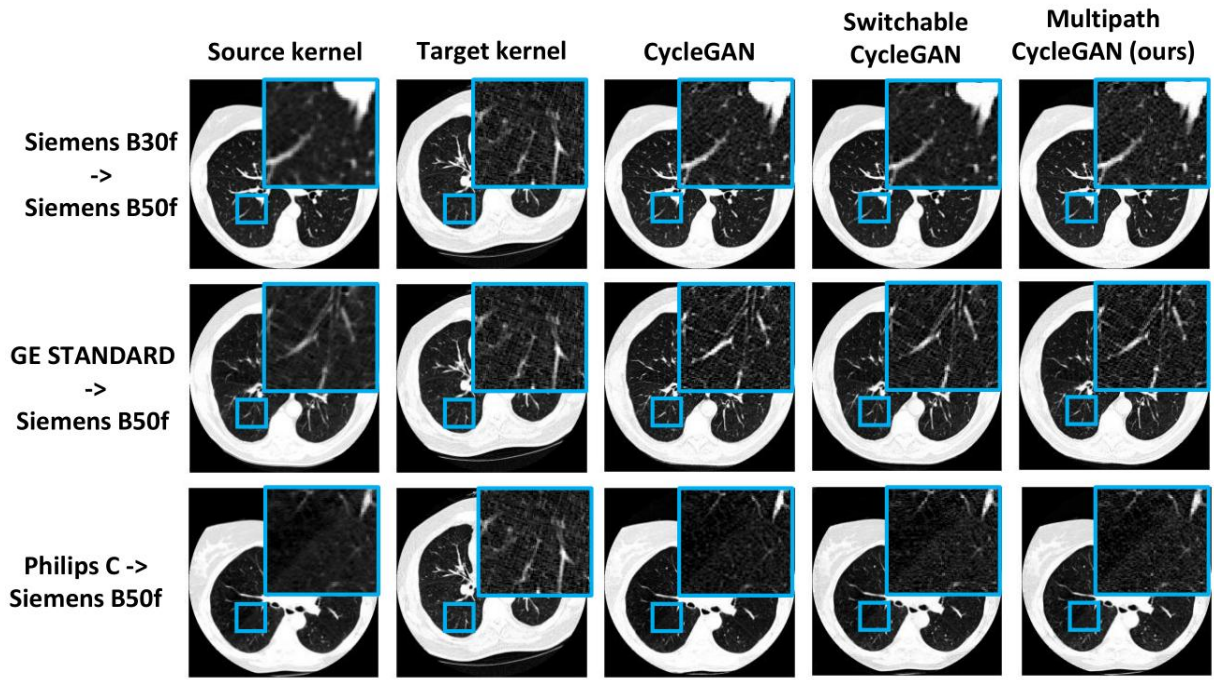

**Figure 8.** Harmonizing a soft kernel to hard kernel is a challenging task since it is difficult to recover high frequency components from a blurry image. We observe that harmonizing all soft kernels to a reference hard kernel restores the high frequency details in the region of interest in the lungs across the baseline models and our proposed approach. However, the switchable cycleGAN shows a smooth texture for the harmonized B50f kernel.

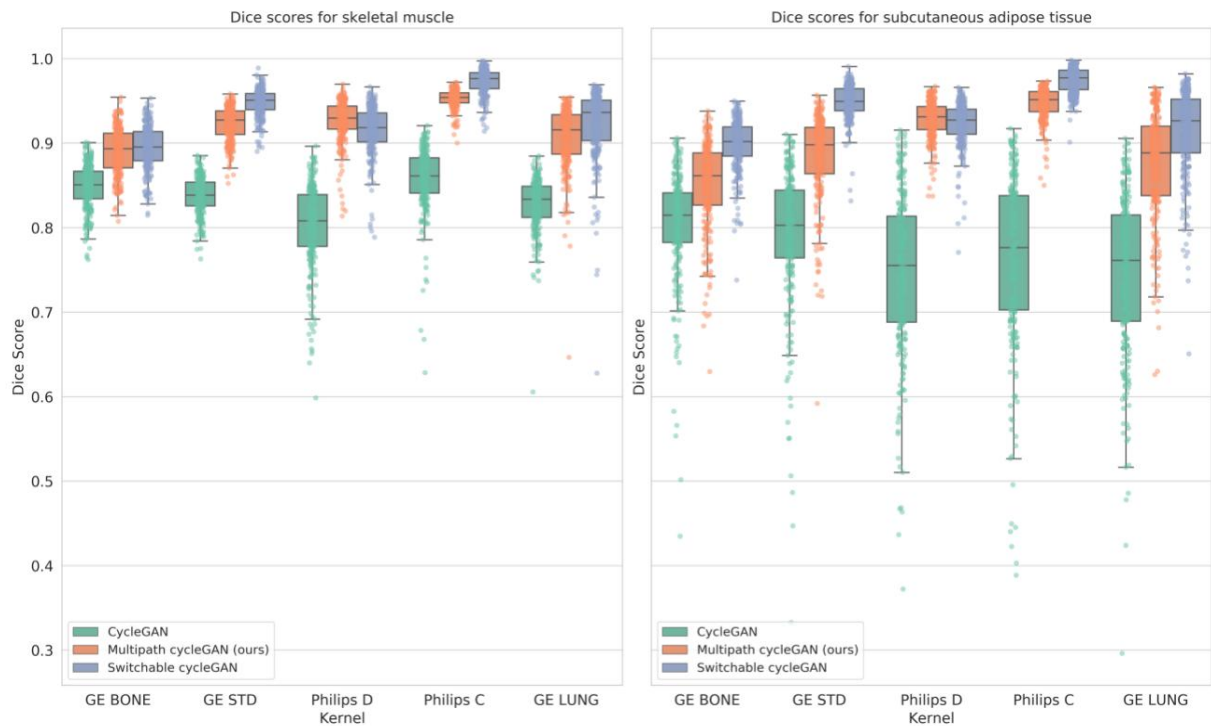

**Figure 9.** We assess the consistency in skeletal muscle and subcutaneous adipose tissue (SAT) before and after harmonization of the unpaired kernels by computing Dice coefficients between the input source kernel and the harmonized target kernel. The multipath cycleGAN preserves anatomy for all kernels as compared to the cycleGAN for muscle and fat. However, when compared to the switchable cycleGAN model, the multipath cycleGAN model either underperforms or achieves similar Dice scores for muscle and SAT.

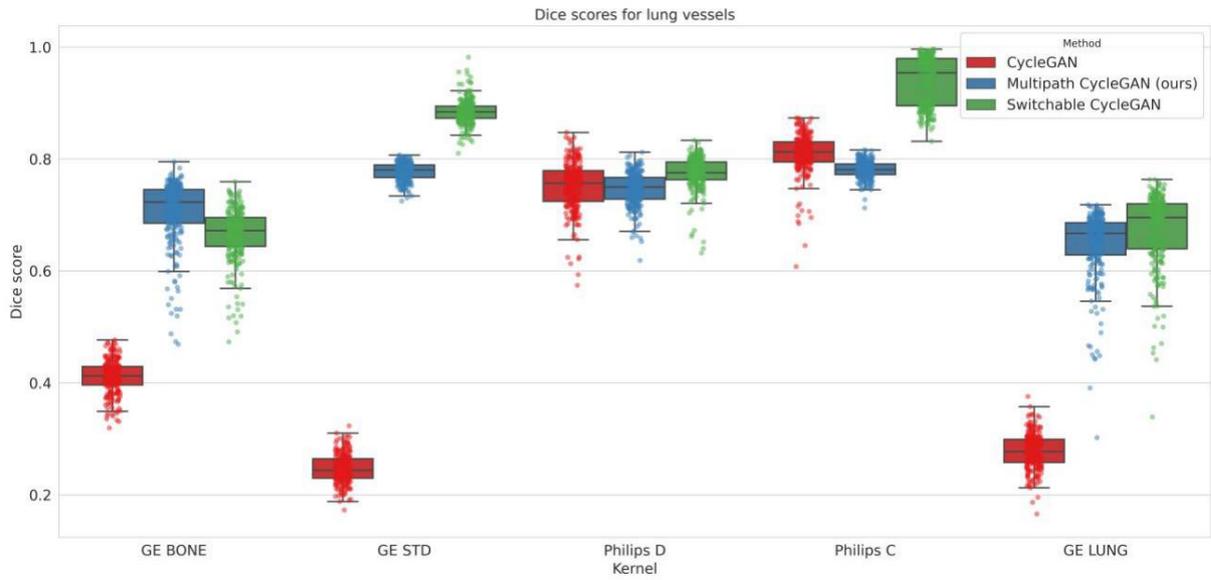

**Figure 10.** We also assess the consistency in lung vessels and airways before and after harmonization on the unpaired kernels using Dice scores computed between the input source kernel and the harmonized target kernel. The proposed multipath cycleGAN shows reasonable overlap when compared to the cycleGAN model for all the GE kernels but underperforms on the Philips kernels. When compared to the switchable cycleGAN model, the multipath cycleGAN does better on the GE BONE kernel while underperforming on other kernels.

**Supplementary Figures**

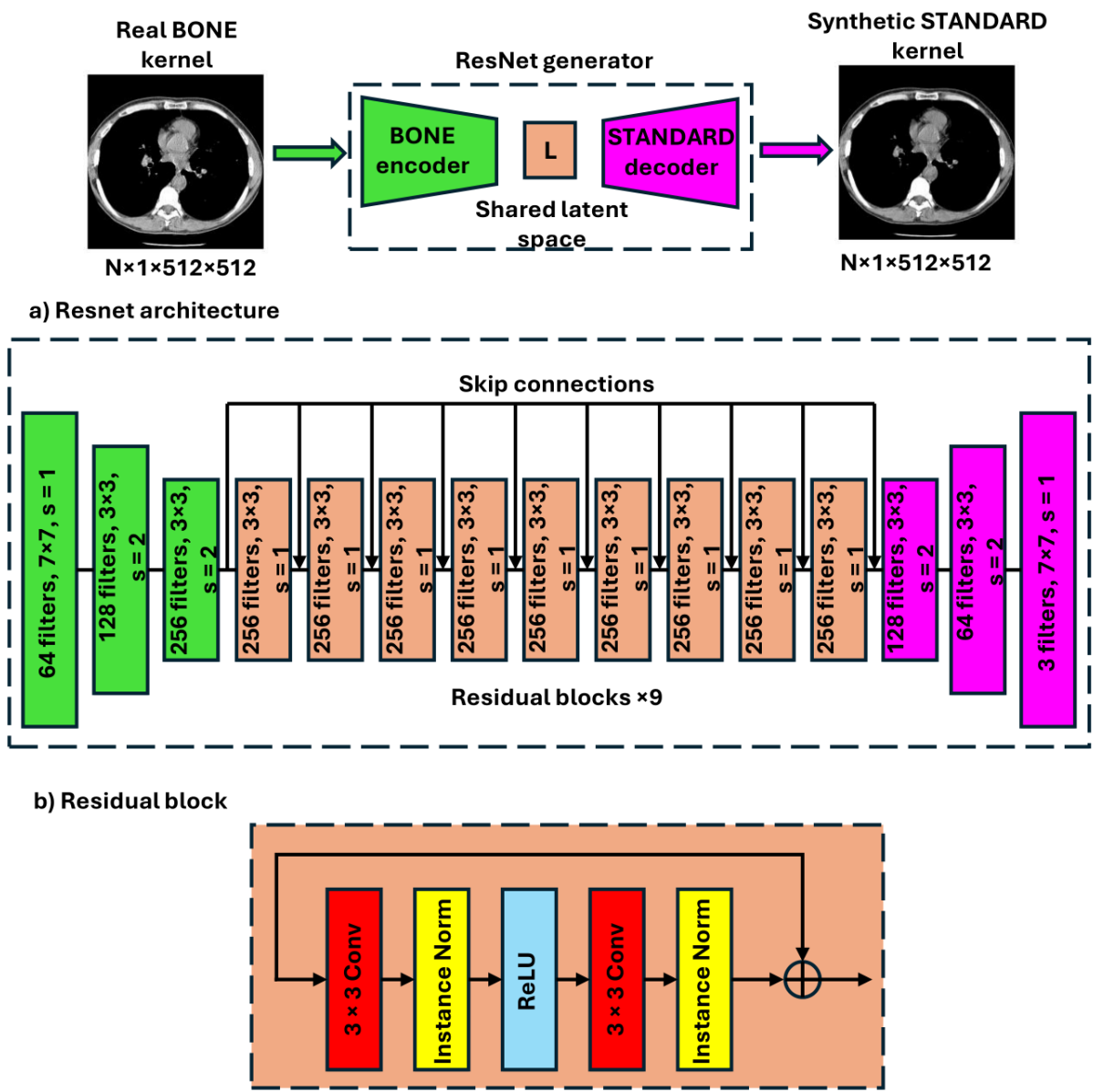

**Figure S1.** The ResNet generator consists of an encoder, shared latent space formed by residual blocks, and a decoder for any given pair of CT reconstruction kernels. **(a)** The encoder consists of three layers with different kernels, filters and strides that progressively downsample the input. The shared latent space is composed of nine residual blocks, each with 256 filters, that preserve the spatial and channel dimensions. The decoder consists of three layers with varying configurations that upsample the feature vector to produce a synthetic image in the target domain. **(b)** Each residual block contains two 3×3 convolution layers with instance normalization and ReLU activation. The input to the first convolution layer is added to the output of the instance normalization layer, resulting in a skip connection.

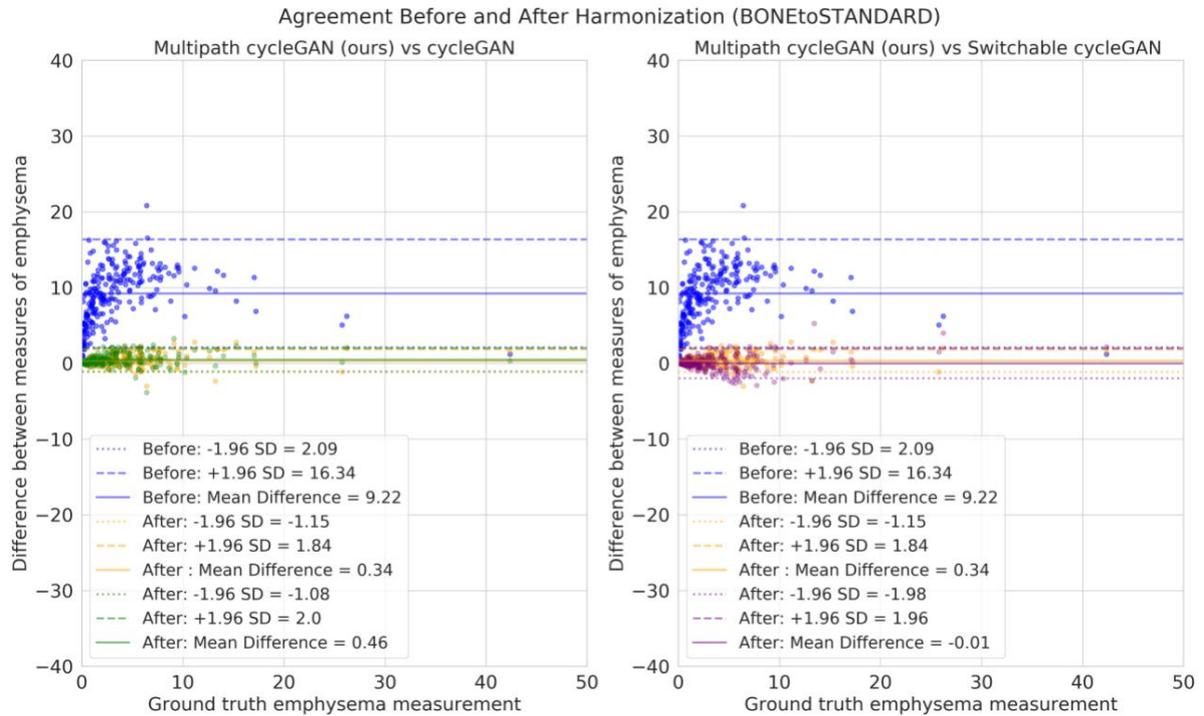

**Figure S2.** Bland Altman style plot for the kernels obtained from the GE manufacturer depicting the performance of multipath cycleGAN versus baseline cycleGAN models. Dashed lines represent confidence intervals, and the solid line represents the mean difference. Blue represents measurements without harmonization, yellow represents the multipath cycleGAN, green represents the standard cycleGAN and purple represents the switchable cycleGAN. Multipath cycleGAN, standard cycleGAN and the switchable cycleGAN model mitigate differences in measurements after harmonization. The multipath cycleGAN achieves a smaller mean difference than the standard cycleGAN while the switchable cycleGAN achieves a mean difference closer to zero.

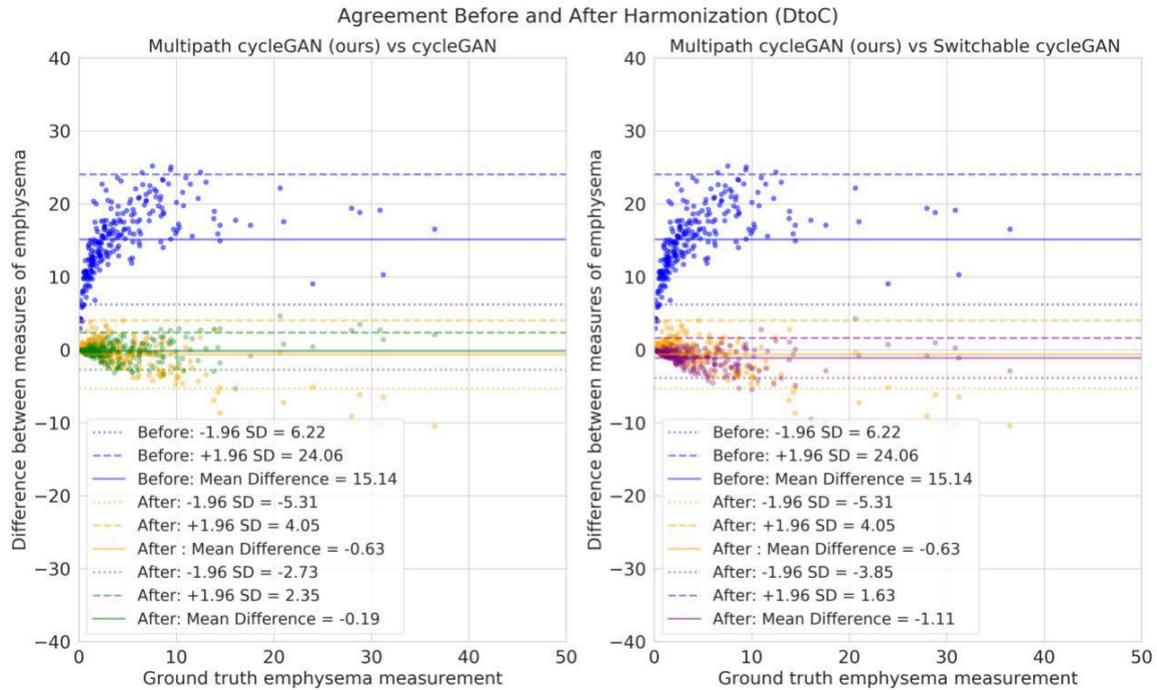

**Figure S3.** Bland Altman style plot for the kernels obtained from the Philips manufacturer depicting the performance of multipath cycleGAN versus baseline cycleGAN models. Dashed lines represent confidence intervals, and the solid line represents the mean difference. Blue represents measurements without harmonization, yellow represents the multipath cycleGAN, green represents the standard cycleGAN and purple represents the switchable cycleGAN. All models mitigate differences in measurements after harmonization. However, the standard cycleGAN and switchable cycleGAN enforce better consistency in measurements compared to the multipath cycleGAN as evidenced by the mean difference and confidence intervals.

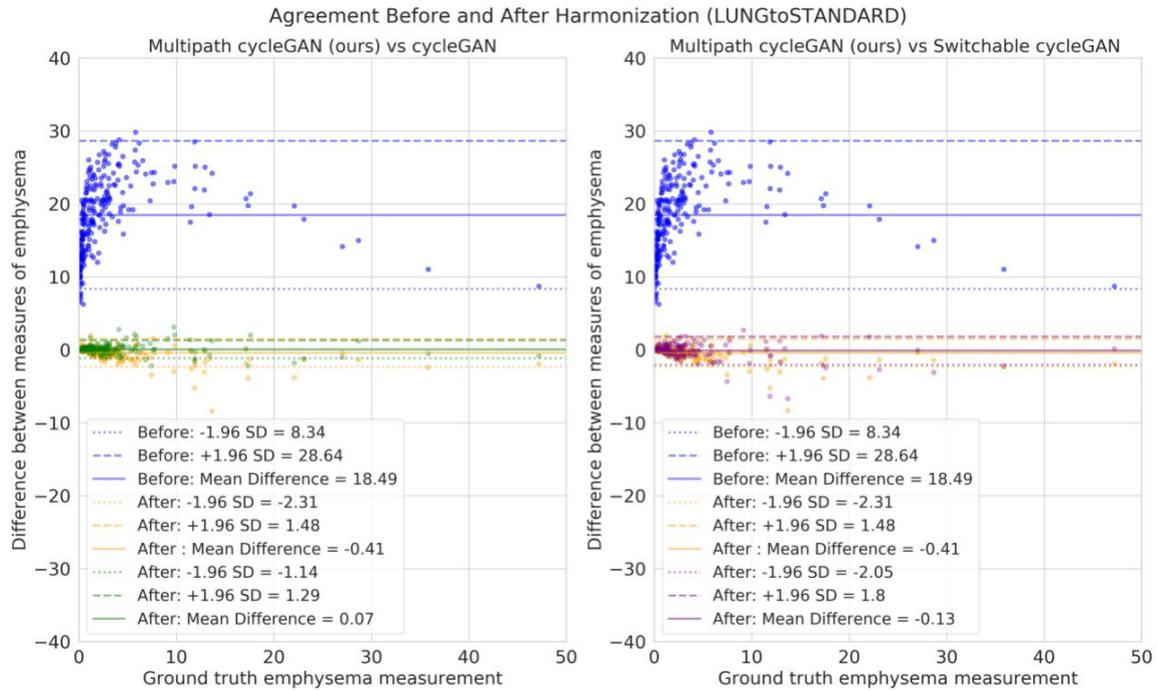

**Figure S4.** Bland Altman style plot for the kernels obtained from the GE manufacturer depicting the performance of multipath cycleGAN versus baseline cycleGAN models. Dashed lines represent confidence intervals, and the solid line represents the mean difference. Blue represents measurements without harmonization, yellow represents the multipath cycleGAN, green represents the standard cycleGAN and purple represents the switchable cycleGAN. Multipath cycleGAN, standard cycleGAN and the switchable cycleGAN model mitigate differences in measurements after harmonization. The standard cycleGAN model is slightly better than the multipath cycleGAN in enforcing consistency in emphysema measurements while the switchable cycleGAN and multipath cycleGAN show comparable performance.

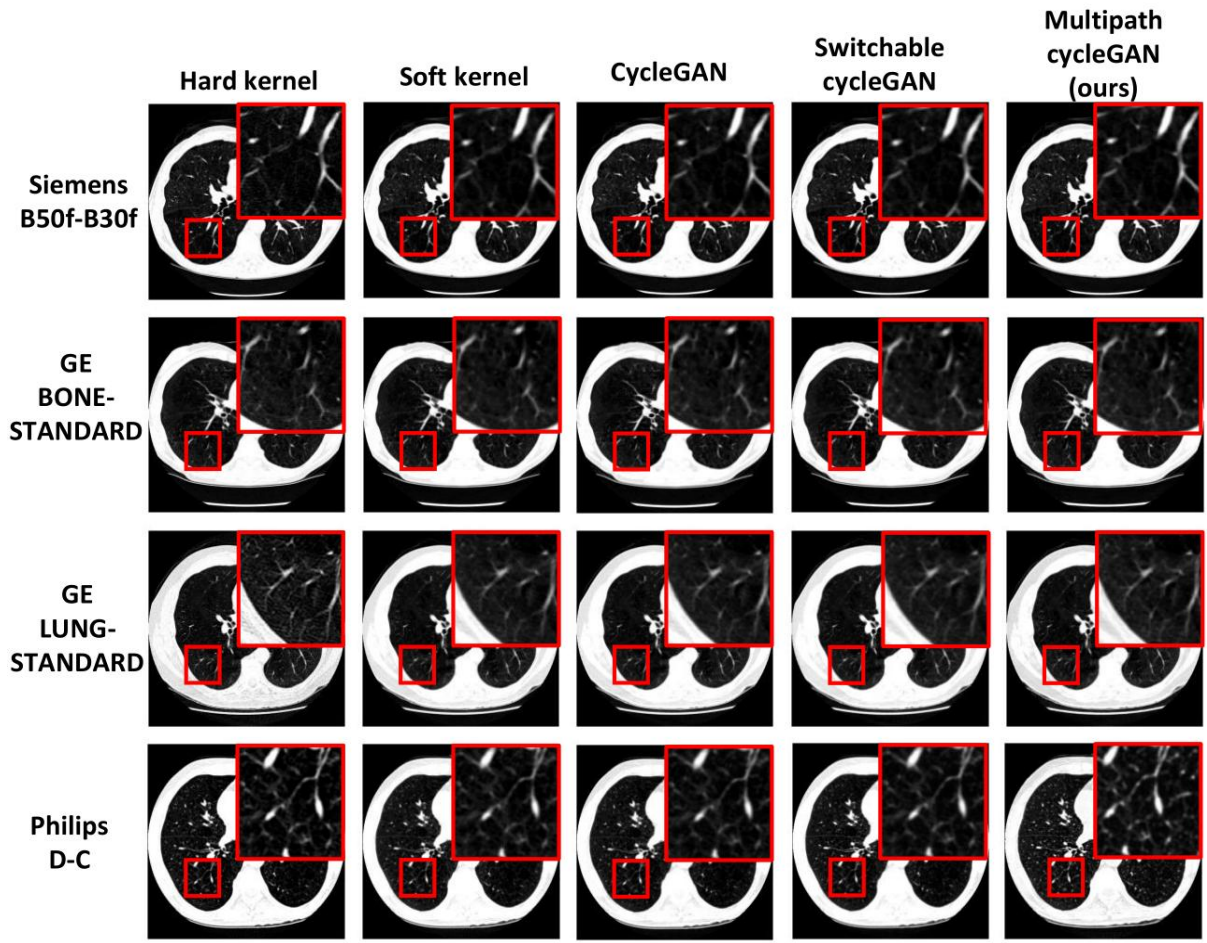

**Figure S5.** We present paired reconstruction kernels before and after harmonization from the 99[th] percentile of subjects showing severe emphysema. The hard and soft kernels exhibit difference in texture of the lung parenchyma which can impact emphysema quantification. Harmonization enforces consistent texture in the lung, ensuring consistent and comparable emphysema assessment.

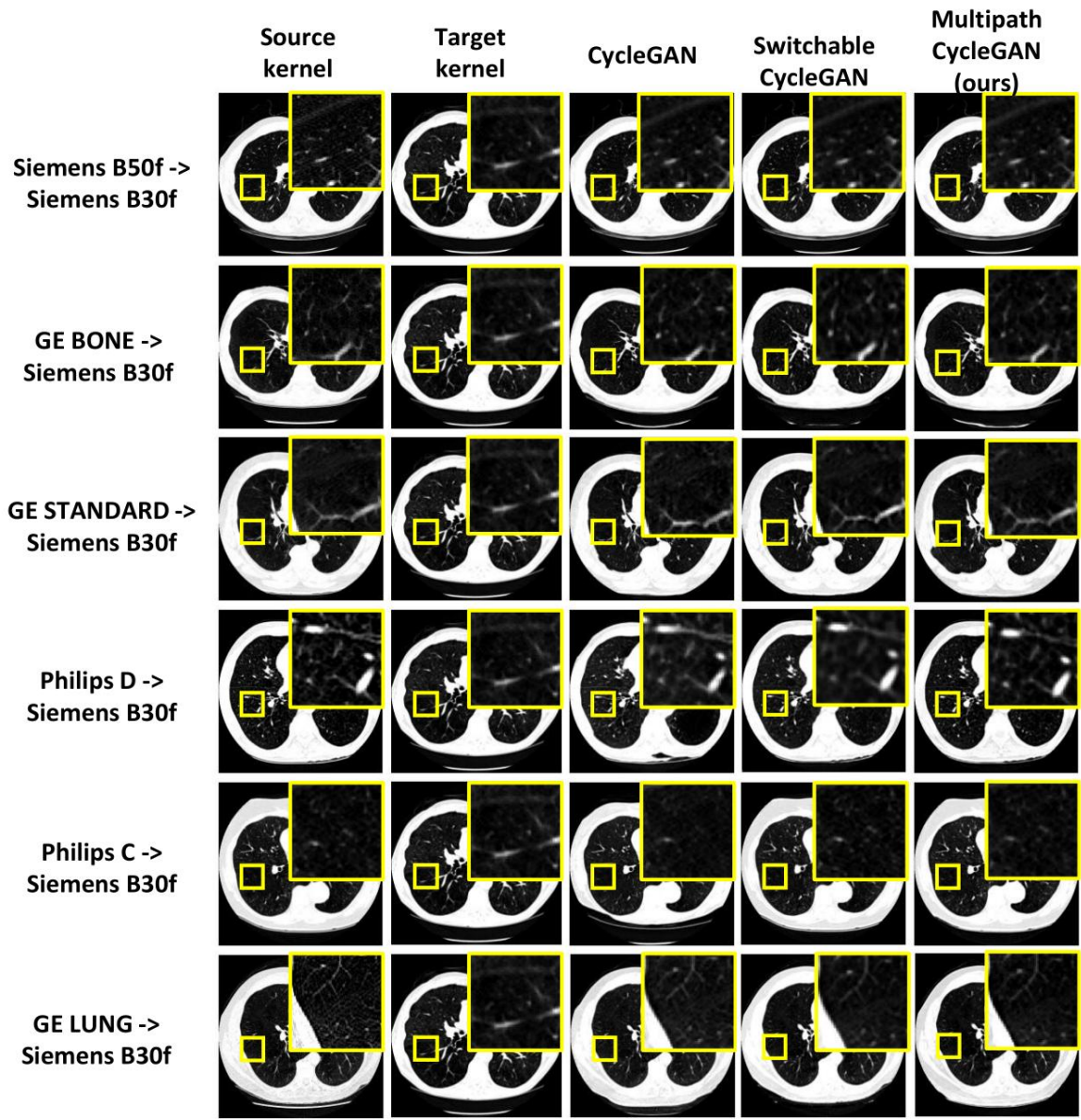

**Figure S6.** We present the subjects from the 99[th] percentile of emphysema distribution for the unpaired reconstruction kernels. Harmonization of all kernels to the reference soft kernel enforces consistent texture in the regions that show emphysema. However, anatomical hallucinations can be seen on the images harmonized by the cycleGAN and the multipath cycleGAN models.

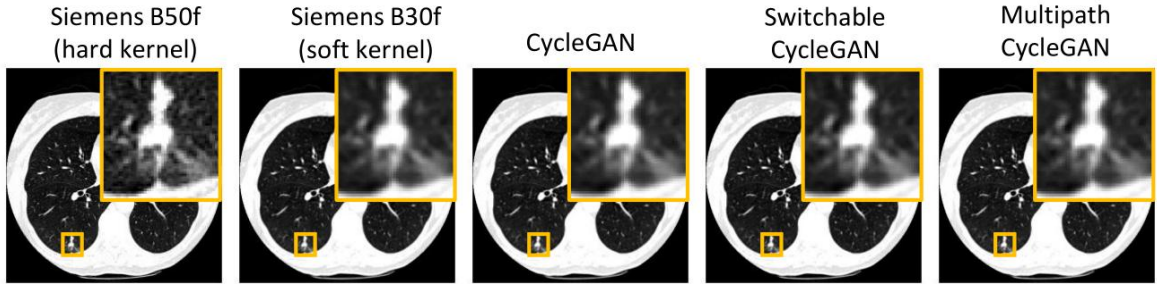

**Figure S7.** The reconstruction kernel impacts the texture of lung nodules. For a given pair of reconstruction kernels from the same subject, the hard kernel (B50f) sharpens the texture of the nodule while the soft kernel (B30f) smoothens it. We observe that harmonization preserves the structural integrity and visibility of the nodule by enforcing consistent texture.

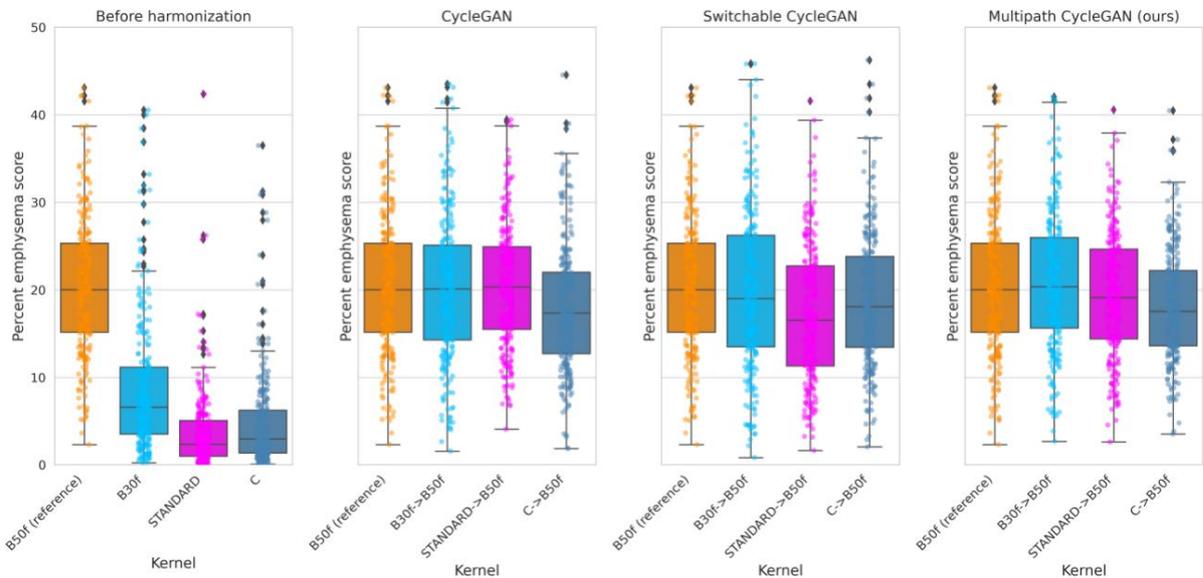

**Figure S8.** Before harmonization, the reference hard kernel, shown as the orange box and strip plot, has a higher range of emphysema scores compared to the soft kernels. Harmonization of all the soft kernels to the reference hard kernel minimizes differences in emphysema measurements We observe that the cycleGAN and our proposed multipath cycleGAN show similar performance with a slight difference in the median. However, the switchable cycleGAN underperforms on all kernels except the Philips C kernel.